\documentclass[preprint]{aastex62}

\usepackage{amsmath}

\usepackage{lipsum}

\usepackage{makecell}
\usepackage{multirow}

\usepackage{nicefrac}

\renewcommand{\arg}[1]{\! \left( #1 \right)}
\newcommand{\pcond}[2]{p \arg{ #1 \, | \, #2 }}

\usepackage{xspace}
\newcommand{\gps}{\ensuremath{g_{\mathrm{P1}}}\xspace}
\newcommand{\rps}{\ensuremath{r_{\mathrm{P1}}}\xspace}
\newcommand{\ips}{\ensuremath{i_{\mathrm{P1}}}\xspace}
\newcommand{\zps}{\ensuremath{z_{\mathrm{P1}}}\xspace}
\newcommand{\yps}{\ensuremath{y_{\mathrm{P1}}}\xspace}

\newcommand{\Egr}{\ensuremath{E \left( g \! - \! r \right)}\xspace}
\newcommand{\EBPRP}{\ensuremath{E \left( BP \! - \! RP \right)}\xspace}

\graphicspath{{./}{figures/}}

\received{?}
\revised{?}
\accepted{?}
\submitjournal{?}

\shorttitle{Bayestar 2019}
\shortauthors{Green et al.}

\begin{document}

\title{A 3D Dust Map Based on Gaia, Pan-STARRS 1 and 2MASS}

\correspondingauthor{Gregory M. Green}
\email{gregorymgreen@gmail.com}

\author[0000-0001-5417-2260]{Gregory M. Green}
\affil{Kavli Institute for Particle Astrophysics and Cosmology, \\
Stanford University \\
452 Lomita Mall, \\
Stanford, CA 94305-4060, USA}

\author[0000-0002-3569-7421]{Edward Schlafly}
\affiliation{Lawrence Berkeley National Laboratory \\ 
One Cyclotron Road \\
Berkeley, CA 94720, USA}
\affiliation{Hubble Fellow}

\author[0000-0002-2250-730X]{Catherine Zucker}
\affiliation{Harvard Astronomy, \\
Harvard-Smithsonian Center for Astrophysics \\
60 Garden St., \\
Cambridge, MA 02138, USA}

\author[0000-0002-5065-9896]{Joshua S. Speagle}
\affiliation{Harvard Astronomy, \\
Harvard-Smithsonian Center for Astrophysics \\
60 Garden St., \\
Cambridge, MA 02138, USA}

\author[0000-0003-2808-275X]{Douglas Finkbeiner}
\affiliation{Harvard Astronomy, \\
Harvard-Smithsonian Center for Astrophysics \\
60 Garden St., \\
Cambridge, MA 02138, USA}



\begin{abstract}

We present a new three-dimensional map of dust reddening, based on Gaia parallaxes and stellar photometry from Pan-STARRS 1 and 2MASS. This map covers the sky north of a declination of $-30^{\circ}$, out to a distance of several kiloparsecs. This new map contains three major improvements over our previous work. First, the inclusion of Gaia parallaxes dramatically improves distance estimates to nearby stars. Second, we incorporate a spatial prior that correlates the dust density across nearby sightlines. This produces a smoother map, with more isotropic clouds and smaller distance uncertainties, particularly to clouds within the nearest kiloparsec. Third, we infer the dust density with a distance resolution that is four times finer than in our previous work, to accommodate the improvements in signal-to-noise enabled by the other improvements. As part of this work, we infer the distances, reddenings and types of 799 million stars. We obtain typical reddening uncertainties that are $\sim$30\% smaller than those reported in the Gaia DR2 catalog, reflecting the greater number of photometric passbands that enter into our analysis. Our 3D dust map can be accessed at \href{https://doi.org/10.7910/DVN/2EJ9TX}{doi:10.7910/DVN/2EJ9TX} or through the Python package \texttt{dustmaps}. Our catalog of stellar parameters can be accessed at \href{https://doi.org/10.7910/DVN/AV9GXO}{doi:10.7910/DVN/AV9GXO}.

\end{abstract}

\keywords{ISM: dust, extinction --- ISM: structure --- Galaxy: structure --- Galaxy: stellar content}


\section{Introduction}
\label{sec:intro}

Dust is both a critical foreground for many astrophysical measurements and a tracer of Galactic structure and star-forming regions. In the ultraviolet, optical and near-infrared, dust causes both extinction and reddening. It is necessary to correct for these effects in order to measure intrinsic luminosities or colors of obscured objects. For extragalactic astronomy, a two-dimensional map of integrated dust extinction and reddening is sufficient, while for sources embedded in the Milky Way, a distance-dependent correction for dust obscuration is necessary. Dust emits strongly in the mid- and far-infrared, and thus provides an important foreground for the Cosmic Microwave Background, further motivating accurate maps of the distribution of Galactic dust.

Far-infrared emission of dust can be used to construct two-dimensional maps of dust reddening. \citet[][hereafter ``SFD'']{SFD98} models far-infrared dust emission as a modified blackbody, deriving maps of dust optical depth and temperature. Assuming a linear relation between dust optical depth at 100~$\mathrm{\mu}$m and reddening at optical wavelengths, SFD calibrates their optical depth map against measurements of $E \arg{B-V}$ from elliptical galaxies, obtaining a full-sky map of Galactic dust reddening at optical wavelengths. Because it is not possible to determine the distance to the emitting dust from far-infrared intensity alone, this method does not recover the three-dimensional distribution of dust. In addition, the relation between far-infrared optical depth and optical reddening depends on the dust composition and grain-size distribution, and thus varies throughout the Galaxy, introducing systematic errors into the resulting reddening maps.

One method of overcoming these limitations is to use optical stellar photometry to trace dust. Stellar distances can be determined (both from spectral models and from geometric parallax measurements), allowing one to trace reddening as a function of both position on the sky and distance. In addition, stellar colors are a more direct measurement of optical reddening, obviating the need to convert from far-infrared optical depth to reddening at optical wavelengths. Several works have now used optical and near-infrared stellar photometry to trace dust in 3D. In lieu of an extensive summary, a few works will be mentioned here. \citet{Marshall2006} uses near-infrared colors of post-main-sequence stars in the Two Micron All Sky Survey \citep[][hereafter ``2MASS'']{Skrutskie2006} to map dust in 3D, while \citet{Berry2012} uses the optical and near-infrared colors of stars in the Sloan Digital Sky Survey \citep{York2000}.

These methods have matured with the introduction of probabilistic models that infer both the distribution of dust and the types and distances of stars. \citet{Sale2014IPHAS} applies a hierarchical Bayesian model of the distribution of dust, as well as stellar distances and types, to the INT Photometric H$\alpha$ Survey of the Northern Galactic Plane \citep[][``IPHAS'']{Drew2005IPHAS}. \citet{Green2015} models Pan-STARRS~1 and 2MASS photometry of $\sim$800 million stars over three-quarters of the sky to create a map that extends to several kiloparsecs. \citet{Green2018} improves upon this map, with both updated Pan-STARRS~1 photometry and methodological improvements. \citet{Lallement2014} uses catalog distances and reddenings of stars to trace dust, but applies a Gaussian process prior to the spatial distribution of dust, enforcing smoothness on small spatial scales.

Recently, the availability of geometric parallax measurements from Gaia \citep{GaiaCollaboration2016Mission} has given a significant boost to the field. Whereas most previous studies of the 3D distribution of dust have relied on stellar spectral energy distribution models to determine distances to stars, Gaia more directly constrains the distances to the hundreds of millions of stars for which it precisely measures parallaxes. In addition, Gaia measures stellar optical photometry, which \citet{Andrae2018DR2Apsis} uses to estimate stellar reddenings and extinctions. Using a method similar to that of \citet{Andrae2018DR2Apsis}, \citet{Chen2018} estimates stellar reddenings from Gaia photometry and parallaxes, producing an all-sky map of dust reddening in 3D. \citet{LeikeEnsslin2019} combines the extinction estimates from \citet{Andrae2018DR2Apsis} with a Gaussian process model for the spatial distribution of dust to map reddening within a few hundred parsecs of the Sun. \citet{Lallement2019} combines Gaia parallaxes with photometry from Gaia and 2MASS to estimate stellar distances and reddenings, and models the spatial distribution of dust as a Gaussian process, obtaining a map that extends to several kiloparsecs in the midplane of the Galaxy.

In the following, we present a new 3D map of dust reddening, based on Gaia parallaxes and broad-band optical and near-infrared photometry. We make several major improvements over \citet{Green2018}, incorporating Gaia parallaxes into our model, improving the distance resolution of the map by a factor of four, and imposing a Gaussian process prior on the distribution of dust. For stars with precisely measured Gaia parallaxes, we obtain dramatically more precise distance determinations than is possible with broad-band photometry alone. The imposition of a Gaussian process prior on the distribution of dust not only regularizes the resulting map, but also allows us to more precisely determine distances to dust clouds within $\sim$1~kpc.

Our method produces not only a map of the 3D distribution of dust reddening, but also a catalog of distances, reddenings and types for 799 million stars. Our 3D dust map may be accessed at \href{https://doi.org/10.7910/DVN/2EJ9TX}{doi:10.7910/DVN/2EJ9TX}, while our catalog of stellar parameters may be accessed at \href{https://doi.org/10.7910/DVN/AV9GXO}{doi:10.7910/DVN/AV9GXO}. In addition, our dust map can be accessed through the Python package \texttt{dustmaps} \citep{Green2018dustmaps}.

The paper is organized as follows. Section \ref{sec:method} reviews our method for inferring stellar properties (Section \ref{sec:single-star}) and the distribution of dust along each sightline (Section \ref{sec:single-sightline}), and then describes a new method for imposing a correlated prior on the dust, linking nearby sightlines (Section \ref{sec:correlated-prior}). This latter process yields a map in which dust densities in nearby voxels, spanning different sightlines, are correlated, dependent on the distance between the voxels. Section \ref{sec:data} describes the survey data we use, from Gaia, Pan-STARRS 1 and 2MASS. Section \ref{sec:results} presents our new 3D dust map, compares it to previous dust maps, assesses evidence in our map for spiral structure in the Milky Way, and presents our catalog of stellar parameters. Section \ref{sec:discussion} discusses ways forward for dust mapping. Finally, we conclude in Section \ref{sec:conclusion}.

\section{Method}
\label{sec:method}

Here, we sketch out our method for determining the three-dimensional distribution of dust, as well as stellar types and distances. In brief, we use stars as tracers of the dust column, modeling the apparent magnitudes and parallaxes of stars as a function of their type, distance and foreground reddening. We then group the stars into discrete sightlines, using the stellar distances and reddenings to constrain the dust as a function of distance.

The method for inferring types and distances for single stars is much the same as in \citet{Green2015} and \citet{Green2018}, with the addition of Gaia parallaxes. However, we have significantly updated our method for sampling the dust reddening along individual sightlines, allowing us to increase the number of distance bins by a factor of four while decreasing the computational runtime, and we have implemented a new iterative scheme for imposing a prior with correlations between nearby sightlines.

\subsection{Single-Star Inference}
\label{sec:single-star}

\begin{deluxetable}{ccccccccc}
    \tablecaption{Extinction vector, $\vec{R}$\label{tab:extinction-vector}}
    \tablehead{
        \multicolumn{5}{c}{Pan-STARRS 1} & \colhead{ } & \multicolumn{3}{c}{2MASS} \\ \cline{1-5} \cline{7-9}
        \colhead{$g$} & \colhead{$r$} & \colhead{$i$} & \colhead{$z$} & \colhead{$y$} & & \colhead{$J$} & \colhead{$H$} & \colhead{$K_s$}
    }
    \startdata
    3.518 & 2.617 & 1.971 & 1.549 & 1.263 & & 0.7927 & 0.4690 & 0.3026 \\
    \enddata
\end{deluxetable}

\begin{deluxetable}{cCcl}
    \tablecaption{Stellar parameters\label{tab:stellar-parameters}}
    \tablehead{
        & \colhead{Parameter} & \colhead{Meaning} & \colhead{Prior}
    }
    \startdata
    & E      & Reddening                        & Flat prior for $E > 0$. \\
    & \mu    & Distance modulus                 & Thin disk, thick disk \& halo components. \\[6pt]
    \multirowcell{2}{$\Theta \, \left\{ \vphantom{ \Bigg| } \right. \hspace{-1.5em}$} & M_r  & $r_{\mathrm{P1}}$-band magnitude & Universal luminosity function. \\
    & \left[ \mathrm{Fe} / \mathrm{H} \right] & Metallicity
    & \makecell[lt]{Gaussian for halo component; \\ Gaussian with mean and variance \\ dependent on height for disk components.} \\
    \enddata
\end{deluxetable}

For an individual star with measured photometry, $\hat{m}$, and parallax, $\hat{\varpi}$, we wish to determine the star's type, $\Theta$, distance modulus, $\mu$, and reddening, $E$. We can write down the posterior probability density of these parameters, given the data, as
\begin{align}
    \pcond{\Theta, \, \mu, \, E}{\hat{m}, \, \hat{\varpi}}
    &= \frac{
        \pcond{\hat{m}, \, \hat{\varpi}}{\Theta, \, \mu, \, E} p \arg{\Theta, \, \mu, \, E}
    }{
        p \arg{\hat{m}, \, \hat{\varpi}}
    } \, .
\end{align}
Since the photometry and parallax are observed, fixed quantities, we can treat $p \arg{\hat{m}, \, \hat{\varpi}}$ as a constant, so
\begin{align}
    \pcond{\Theta, \, \mu, \, E}{\hat{m}, \, \hat{\varpi}}
    \propto
    \pcond{\hat{m}, \, \hat{\varpi}}{\Theta, \, \mu, \, E} p \arg{\Theta, \, \mu, \, E} \, .
    \label{eqn:single-star-posterior}
\end{align}
If we are given the type, distance modulus and reddening of a star, we can look up its model magnitudes, $M_{\mathrm{model}} \arg{\Theta}$, using the stellar templates described in \citet{Green2015}, and predict its parallax, which is simply a function of distance modulus, $\varpi \arg{\mu}$. Comparison of these model observables with the observed photometry and parallax yields the likelihood term, $\pcond{\hat{m}, \, \hat{\varpi}}{\Theta, \, \mu, \, E}$, which is a multivariate Gaussian:
\begin{align}
    \hat{m}
    &\sim
    \mathcal{N} \left[ \,
        M_{\mathrm{model}} \arg{\Theta} + \mu + A \arg{E}, \,
        \sigma_m
    \, \right]
    \, , \\
    \hat{\varpi}
    &\sim
    \mathcal{N} \left[ \,
        \varpi \arg{\mu} , \,
        \sigma_{\varpi}
    \, \right]
    \, .
\end{align}
where $\sigma_m$ and $\sigma_{\varpi}$ are the uncertainty in the observed magnitudes and parallax, and $A \arg{E}$ is the extinction in the observed passbands, assumed to be a linear function of reddening, $E$. We assume that for each star, extinction is given by
\begin{align}
    \vec{A} \arg{E} = E \, \vec{R} \, ,
\end{align}
where $\vec{R}$ is the ``extinction vector,'' relating a scalar reddening to the extinction in each passband. We use the $R_V = 3.3$ empirical reddening vector from \citet{Schlafly2016ExtinctionVector}, converting it to an extinction vector by requiring that $A_H / A_K = 1.55$ \citep{Indebetouw2005} and that $E \left( g_{\mathrm{P1}} - r_{\mathrm{P1}} \right) = 0.901 \, \mathrm{mag}$ when $E = 1 \, \mathrm{mag}$. The latter choice puts our measure of reddening on a similar scale as SFD \citep{SchlaflyFinkbeiner2011}. The resulting extinction vector is given in Table \ref{tab:extinction-vector}. This extinction vector is slightly different from the one we use in \citet{Green2018}, as we normalize the vector in that work by setting extinction in the WISE W2 band to zero. The difference between the two extinction vectors is most pronounced in the 2MASS passbands.

We place priors on the distribution of stars of different types throughout the Galaxy, yielding a joint prior on distance modulus and type, $p \arg{\Theta, \, \mu}$. These priors are described in \citet{Green2014}, and are summarized in Table \ref{tab:stellar-parameters}. For reasons that will become clear in Section \ref{sec:single-sightline}, we place a flat prior on the reddening of each star: $p \arg{E} = \mathrm{const}$.

We evaluate the posterior density of each star, $\pcond{\mu, \, E}{\hat{m}, \, \hat{\varpi}}$, on a grid of distance modulus $\mu \in \left[ 4 \, \mathrm{mag}, \, 19 \, \mathrm{mag} \right]$ in steps of 0.125~mag and $E \in \left[ 0 \, \mathrm{mag}, \, 7 \, \mathrm{mag} \right]$ in steps of 0.01~mag. Note that stellar type, $\Theta$, has been integrated out here. Our procedure for evaluating this posterior density quickly over a regular grid is explained in Appendix \ref{app:grid-stars}. This new method is both faster and more accurate than the evaluation technique used in our previous work, which involved kernel density estimation on samples drawn from an MCMC chain.

As in \citet{Green2015} and \citet{Green2018}, we smooth the stellar probability density functions along the reddening axis with a Gaussian kernel, which is equivalent to treating reddening within each pixel as a white noise process at any given distance. This takes into account the small-scale power in the dust density spectrum, allowing reddening to vary across a single angular pixel. The Gaussian smoothing applied is a fraction of the reddening in the pixel, as explained in Section 2.2 of \citet{Green2015}. We make one change to the smoothing method developed in \citet{Green2015}, in that we increase the minimum reddening smoothing percentage to 15\% and the maximum smoothing percentage to 50\%.

Not all of the point sources that we model are indeed stars -- our sample is inevitably contaminated by compact galaxies and quasars. We attempt to filter out galaxies by enforcing a compactness criterion on point sources (see Section \ref{sec:data}), but at the faint end of our source catalog, this criterion becomes less effective, both because morphology of sources becomes more difficult to determine and because galaxies become more numerous in relation to stars. There are additionally certain stellar types, such as O and B stars, which are not included in our stellar template library. In order to filter out such sources, we make a goodness-of-fit cut after modeling the stellar parameters. In detail, we filter out stars for which the minimum $\chi^2 / \mathrm{passband} > 5$ (i.e., for which no stellar template accurately models the observed photometry). For the purposes of this cut, we count stellar parallax as an additional photometric band. In previous work, we cut on the Bayesian evidence of the model for each individual star. Our new cut is similar in spirit, but does not depend on the choice of priors, relying instead only on the likelihood. Approximately 1.4\% of point sources fail this goodness-of-fit cut, and are therefore not used to determine line-of-sight reddening.

\subsection{Single-Sightline Inference}
\label{sec:single-sightline}

In this section, we will describe how to use the stars within one individual pixel to infer the dust density as a function of distance. The way in which we do this for one isolated sightline is described in detail in \citet{Green2018}, but we will recap it here.

We split the sightline up into discrete distance bins, and parameterize the logarithm of the increase in reddening in the distance bins by $\alpha$. We want to know what distribution of dust along the line of sight and what choice of stellar types and distances is consistent with the observed stellar magnitudes and parallaxes. The posterior on the line-of-sight dust distribution is given by
\begin{align}
    \pcond{\alpha}{\!\left\{ \hat{m}, \, \hat{\varpi} \right\}}
    &= \frac{1}{\mathcal{Z}} \,
    \pcond{\left\{ \hat{m}, \, \hat{\varpi} \right\}\!}{\alpha}
    p \arg{\alpha}
    \, ,
\end{align}
where $\left\{ \hat{m}, \, \hat{\varpi} \right\}$ denotes the photometry and parallaxes for all the stars in the sightline, and $\mathcal{Z} = p \arg{\left\{ \hat{m}, \, \hat{\varpi} \right\}}$ is a constant for any given observed data. Assuming that the stars are independent, we get
\begin{align}
    \pcond{\alpha}{\!\left\{ \hat{m}, \, \hat{\varpi} \right\}}
    &= \frac{1}{\mathcal{Z}} \,
    p \arg{\alpha}
    \prod_{i \, \in \, \mathrm{stars}}
    \pcond{\hat{m}_i, \, \hat{\varpi}_i}{\alpha}
    \, .
    \label{eqn:los-posterior-intermediate}
\end{align}
The term $p \arg{\alpha}$ encodes our prior expectations about the amount of dust in each distance bin along the sightline. As in \citet{Green2018} and \citet{Green2015}, we put a log-normal prior on the jump in reddening in each distance bin, based on a simple, smooth model of the distribution of dust in the Milky Way. In Section \ref{sec:correlated-prior}, we describe how to impose a prior that links the dust in nearby sightlines.

We have a term $\pcond{\hat{m}_i, \, \hat{\varpi}_i}{\alpha}$ for each star $i$. In what follows, we'll drop the subscript $i$, and it is assumed we are discussing a single star. We can expand this term as
\begin{align}
    \pcond{\hat{m}, \, \hat{\varpi}}{\alpha}
    &= \int
    \pcond{\hat{m}, \, \hat{\varpi}, \, \Theta, \, \mu}{\alpha}
    \, \mathrm{d}\Theta \, \mathrm{d}\mu
    \\
    &= \int
    \pcond{\hat{m}, \, \hat{\varpi}}{\Theta, \, \mu, \, \alpha}
    p \arg{\Theta , \, \mu}
    \, \mathrm{d}\Theta \, \mathrm{d}\mu \, .
    \label{eqn:star-double-integral}
\end{align}
Note the similarity between the term in the integrand and the right-hand side of Eq. \eqref{eqn:single-star-posterior}. In fact, these two terms are exactly identical. The observed stellar photometry and parallax are dependent only on the stellar type, distance and reddening. Given a line-of-sight distribution of reddening, $\alpha$, and stellar distance modulus, $\mu$, we can calculate the stellar reddening, $E$. Thus,
\begin{align}
    \pcond{\hat{m}, \, \hat{\varpi}}{\Theta, \, \mu, \, \alpha}
    &= \pcond{\hat{m}, \, \hat{\varpi}}{\Theta, \, \mu, \, E} \, ,
\end{align}
where $E = E \arg{\alpha, \, \mu}$. Recall from Section \ref{sec:single-star} that we placed a uniform prior on the stellar reddening, $E$. As a consequence,
\begin{align}
    p \arg{\Theta , \, \mu} \propto p \arg{\Theta , \, \mu, \, E} \, .
\end{align}
We define the function
\begin{align}
    \tilde{p} \arg{\mu, \, E}
    &\equiv
    \int
    \pcond{\hat{m}, \, \hat{\varpi}}{\Theta, \, \mu, \, E}
    p \arg{\Theta, \, \mu}
    \, \mathrm{d}\Theta \, ,
\end{align}
which is pre-computed for each star on a grid in $\mu$ and $E$, according to the method laid out in Appendix \ref{app:grid-stars}. In terms of this function, Eq. \eqref{eqn:star-double-integral} becomes
\begin{align}
    \pcond{\hat{m}, \, \hat{\varpi}}{\alpha}
    &= \int
    \tilde{p} \arg{\mu, \, E \arg{\alpha, \mu}}
    \, \mathrm{d}\mu \, .
\end{align}
Plugging this all into Eq. \eqref{eqn:los-posterior-intermediate}, we obtain
\begin{align}
    \pcond{\alpha}{\!\left\{ \hat{m}, \, \hat{\varpi} \right\}}
    &\propto
    p \arg{ \alpha}
    \prod_{i \, \in \, \mathrm{stars}}
    \int
    \tilde{p}_i \arg{\mu_i, \, E \arg{\alpha, \mu_i}}
    \, \mathrm{d}\mu_i
    \, .
    \label{eqn:los-posterior-final}
\end{align}
Conceptually, the above means that the posterior on the line-of-sight reddening is a product of the prior and a line integral tracing the distance-reddening curve through each individual stellar posterior. This expression is generally quick to compute, once the pre-computed stellar posteriors, $\tilde{p}_i \arg{\mu, \, E}$, are available.

We model the line-of-sight reddening as a step function, with $\alpha$ encoding the increase in reddening in each distance bin. The jumps in reddening are discretized, as integer multiples of 0.01~mag. The distance bins are spaced evenly in distance modulus, from $\mu = 4~\mathrm{mag}$ to 19~mag, with a bin spacing of 0.125~mag. Relative to our previous work \citep{Green2014,Green2015,Green2018}, we find that our discretized parameterization of reddening significantly speeds up the evaluation of the line integrals in Eq. \eqref{eqn:los-posterior-final}, as well as the convergence of our MCMC sampler.

\subsection{Correlated prior}
\label{sec:correlated-prior}

In the previous section, we showed how to determine the distance-reddening relation along one isolated sightline. However, 3D dust maps created sightline-by-sightline display a number of undesirable and unphysical features.

Because distance is fundamentally more difficult to determine than angular position on the sky, dust clouds in 3D dust maps constructed from individual sightlines are typically highly elongated in the radial direction. This is an unphysical feature, which violates the Copernican principle. We might reasonably expect, \textit{a priori}, for dust clouds to be oriented randomly with respect to the Sun (or at least, to not be oriented in some special configuration with respect to the Sun).

Our technique for measuring the distribution of dust along the line of sight depends on bracketing the dust with foreground and background stars. Because of the stochastic nature of the distribution of stars throughout the Galaxy, a particular dust cloud may be tightly constrained in distance by foreground and background stars in one sightline, but poorly constrained in distance in a neighboring sightline. This leads to 3D dust maps in which the precision of our distance estimate for the same cloud can vary significantly across different sightlines. \textit{A priori}, we expect the dust density field to be smooth on small scales, so that if a dust cloud appears at some distance in one sightline, it is likely to appear at a similar distance in neighboring sightlines.

In order to solve these problems, several works have suggested placing Gaussian process priors on either the dust reddening density (i.e., the derivative of reddening with distance) \citep{Rezaei2017Method}, or on the logarithm of the dust reddening density \citep{Lallement2014,LeikeEnsslin2019}. There are advantages and disadvantages to each approach. If dust reddening density is a Gaussian process, then line integrals through the reddening density are themselves Gaussian, meaning that the integrated reddenings to observed stars are a Gaussian process. This feature is exploited by \citet{Rezaei2018Spirals} to avoid having to explicitly model the dust reddening density throughout space, but rather only the integrated reddenings to observed stars. However, dust reddening density cannot be negative and varies over several orders of magnitude, from the diffuse interstellar medium to dense molecular cores. For these reasons, modeling the logarithm of the dust density as a Gaussian process is more physically natural, and is the approach taken in this paper. Both approaches encode the expected smoothness of the dust density field, and will tend to produce maps with round clouds, not stretched along the line of sight. The true interstellar medium is not a Gaussian process, as it contains filamentary structure that is not fully described by a two-point correlation function, yet one may reasonably expect that a Gaussian process prior that correlates dust density in nearby sightlines will produce more physical results than a completely uncorrelated prior in which neighboring sightlines are independent of one another. A third possibility which is worth mentioning is to model the logarithm of the \textit{integrated} extinction to each point in space as a Gaussian process \citep{SaleMagorrian2014-GP}. This has the advantage of being less computationally taxing than modeling the logarithm of the dust extinction density as a Gaussian process, but is somewhat less physical, as it does not encode our knowledge that integrated extinction increases with distance, and even allows negative extinction density.

In \citet{Green2015} and \citet{Green2018}, we modeled the logarithm of the reddening density in each voxel (one distance bin in one sightline) as a Gaussian, with a mean given by the smooth disk component of the model of the Galaxy from \citet{DrimmelSpergel2001}, and a fixed variance in each bin. In this paper, we alter this prior by introducing covariances between the voxels, including covariances between voxels in different sightlines:
\begin{align}
    \alpha_1, \alpha_2, \ldots, \alpha_N
    \sim
    \mathcal{N} \arg{
        \bar{\alpha}_1, \bar{\alpha}_2, \ldots, \bar{\alpha}_N; \,
        K
    }
    \, ,
\end{align}
where $\alpha_k$ is the logarithm of the reddening density in a single voxel, $N$ is the total number of voxels (in multiple sightlines), $\bar{\alpha}_k$ is the expectation of the logarithm of the reddening density in voxel $k$, given by a smooth Galactic model (the same as used in \citealp{Green2018}), and $K$ is a covariance matrix. The covariance between any two voxels is a function of the distance between their centers -- this function is called the ``kernel.'' The kernel contains information about the power spectrum of the interstellar medium, and in principle could be derived from physical considerations (see, e.g., \citealp{SaleMagorrian2014-GP}, which assumes Kolmogorov turbulence in the interstellar medium), or parameterized and inferred itself (as in \citealp{LeikeEnsslin2019}). In this work, we choose instead to fix the kernel, which reduces the computational complexity of the problem.

In Section \ref{sec:single-sightline}, we worked out the posterior density on reddening in a single sightline. With our new Gaussian process prior, the joint posterior density of reddening in many sightlines is similar to Eq. \eqref{eqn:los-posterior-final}:
\begin{align}
    \pcond{\left\{ \alpha \right\}}{\!\left\{ \hat{m}, \, \hat{\varpi} \right\}}
    & \, \propto \,
    p \arg{ \alpha_1, \, \ldots, \, \alpha_N} \!\!\!\!\!
    \prod_{n \, \in \, \mathrm{sightlines}}
    \left[ \,
    \prod_{i \, \in \, \mathrm{stars\, in\, } n}
    \int
    \tilde{p}_i \arg{\mu_i, \, E \arg{\alpha_n, \mu_i}}
    \, \mathrm{d}\mu_i
    \, \right]
    \, .
    \label{eqn:los-posterior-joint}
\end{align}
The joint prior, $p \arg{ \alpha_1, \, \ldots, \, \alpha_N}$, is the Gaussian process prior over the logarithm of the reddening density in all the voxels in the entire volume under consideration, which spans multiple sightlines.

\subsubsection{Covariance kernel}

In this section, we describe the kernel, which sets the covariance between the logarithm of the reddening density in any two voxels. We choose a modified exponential kernel, for which covariance falls exponentially at large distances, but which smoothly asymptotes to a maximum value within distances $r \lesssim r_0$. This effective inner radius limits the maximum covariance between any two voxels, which helps to make the problem more computationally tractable. Let $\alpha_k$ and $\alpha_{k^{\prime}}$ be the logarithm of the dust density in two different voxels, whose centers lie at $\vec{r}_k$ and $\vec{r}_{k^{\prime}}$, respectively, and $r_{k k^{\prime}}$ be the physical distance between the centers of the voxels. Our kernel is given by
\begin{align}
    K \arg{ \vec{r}_k , \, \vec{r}_{k^{\prime}}}
    = \sigma_0^2 \, A \, \exp \left[
      -\frac{1}{\ell} \left( r_{k k^{\prime}}^{\gamma} + r_0^{\gamma}
        \right)^{\nicefrac{1}{\gamma}}
        + \frac{r_0}{\ell}
    \right] \, ,
    \label{eqn:kernel}
\end{align}
where $\ell$ is a distance scale over which the correlation between voxels decays, $A$ is the asymptotic value of the correlation coefficient for small physical separations, $r_0$ is the inner radius at which the correlation begins to smoothly asymptotes to $A$, and $\gamma$ defines how smoothly the cap is imposed. We set the variance in the logarithmic reddening density (i.e., the diagonal of the covariance matrix) to $\sigma_0^2$. This parameter thus sets the overall scale of variation in reddening from the smooth Galactic model. For $\vec{r}_k = \vec{r}_{k^{\prime}}$ (zero distance, representing the variance in a single voxel), the kernel is defined to be $\sigma_0^2$. We set $\sigma_0 = 1.6$, which yields a prior with approximately the same mean and variance in the expected reddening per unit distance as the prior chosen in \citet{Green2018}, taking into account the increased distance resolution of our new map.

\begin{sloppypar} 
For the remaining kernel parameters, we choose ${\ell = 1.5 \, \mathrm{pc}}$, ${r_0 = 0.75 \, \mathrm{pc}}$, ${\gamma = 4}$ and ${A = \exp \arg{-\nicefrac{1}{2}} \approx 0.6}$. The scale length, $\ell$, is chosen to be similar to the transverse distance between neighboring sightlines, with a typical angular separation of $6.8^{\prime}$, at a radial distance of 1~kpc, which comes to ${\sim \! 2 \, \mathrm{pc}}$. The inner radius, $r_0$, is set so that the prior is not dominated by correlations between the nearest distance bins, which are separated by smaller transverse distances. Although in theory, large correlations between nearby voxels are not a problem, they can pose difficulties for the iterative approach we develop in Section \ref{sec:iterative-scheme}.
\end{sloppypar}

We give a summary of our chosen kernel parameters in Table \ref{tab:cov-kernel}.

\subsubsection{Importance sampling}

Our goal is to impose a correlated prior that links dust reddening densities across different sightlines. In order to do this, one naively has to infer the dust reddening density in all voxels simultaneously. A naive approach would involve calculating the covariance matrix, $K$, between all of the voxels in the map, inverting it, and then using MCMC to sample the posterior density (Eq. \ref{eqn:los-posterior-joint}) of the logarithmic reddening densities in all the voxels. With millions of sightlines, each with $\sim$100 distance bins, this would require us to invert an enormous covariance matrix, and then to sample from a space with hundreds of millions of parameters, a daunting task.

One way of reducing this problem to a feasible one, which we employ in this work, is to infer the dust reddening density only on restricted patches of sky, rather than over the entire sky at once. This limits the size of the covariance matrix which must be calculated and inverted, and also decreases the dimensionality of the parameter space which must be sampled. Rather than sampling from the reddening density over all voxels in the entire map simultaneously, we instead analyze small patches of sky independently. A second technique one might try is importance sampling, which we will flesh out here.

In importance sampling, one wishes to sample from a distribution $p \arg{\theta}$, which might be expensive to compute. One begins instead by drawing points from a distribution which is easier to sample, which we will call $p^{\prime} \arg{\theta}$. One then reweights each sample $j$ by the ratio
\begin{align}
    w_j = \frac{p \arg{\theta_j}}{p^{\prime} \arg{\theta_j}} \, ,
    \label{eqn:importance-weight}
\end{align}
termed the ``importance weight.'' The weighted samples, $\left\{ \left( \theta_j, \, w_j \right) \right\}$, are then representative of the distribution $p \arg{\theta}$. Importance sampling works well when the distribution $p^{\prime} \arg{\theta}$ is a good approximation to the true distribution, $p \arg{\theta}$. In such cases, the weights, $\left\{ w \right\}$, will be close to unity. If $p^{\prime} \arg{\theta}$ is a poor approximation to $p \arg{\theta}$, then the samples, $\left\{ \theta \right\}$, will not be concentrated in the regions of high probability density $p \arg{\theta}$ and will therefore receive near-zero weights, while the few samples which happen to lie in high-probability-density regions will receive large weights.

In our case, we have a Gaussian process prior on the logarithm of the reddening density in each voxel $k$, $\alpha_k$:
\begin{align}
    p \arg{\alpha_0, \, \alpha_1, \, \ldots, \, \alpha_N}
    \, ,
\end{align}
where $N$ is the number of voxels. If the correlations between voxels are small, we can approximate this distribution as a product of independent distributions:
\begin{align}
    p^{\prime} \arg{\alpha_0, \, \alpha_1, \, \ldots, \, \alpha_N}
    = p \arg{\alpha_0} p \arg{\alpha_1} \cdots p \arg{\alpha_N} \, .
\end{align}
We can sample from this uncorrelated, approximate distribution, and then reweight the resulting samples using the ratio
\begin{align}
    w = \frac{
        p \arg{\alpha_0, \, \alpha_1, \, \ldots, \, \alpha_N}
    }{
        p \arg{\alpha_0} p \arg{\alpha_1} \cdots p \arg{\alpha_N}
    } \, .
\end{align}
This method is only effective if the correlations between voxels are small, making the uncorrelated case a good approximation to the correlated case. However, in our case, the uncorrelated prior is sufficiently different from our desired prior that importance sampling would be inefficient.

\subsubsection{An iterative scheme for imposing the correlated prior}
\label{sec:iterative-scheme}

Our central idea here will be to first sample from the uncorrelated case, and then over the course of several iterations, to resample the individual pixels in a way that progressively yields a better approximation to our desired Gaussian process prior.

Our iterative scheme begins by sampling from each sightline individually. For each sightline, $n$, we obtain samples of $\alpha_n$, the logarithm of the dust reddening density in all the distance bins along the line of sight. We then iteratively update each sightline.

In order to update one sightline, we select a small patch of sky surrounding it. We will call the reddening density in the central sightline $\alpha_n$, and the reddening density in the neighboring sightlines $\alpha_{\backslash n}$. We resample the entire patch of sky, but treat the central sightline differently from the neighbors. The reddening profile in each neighbor is represented by a sample from the previous iteration, while the reddenings along central sightline are allowed to take on any value. The neighboring sightlines need only be resampled in order to impose the correct prior on the central sightline, and we only store samples of the reddening density in the central sightline.

We begin by initializing the reddening density in each neighboring sightline to a random sample from the previous iteration, and set the reddening density in the central sightline to an initial guess. We then alternate between updating the central and the neighboring sightlines.

To update one sightline, holding all the others fixed, we need to sample from the conditional distribution
\begin{align}
    \pcond{\alpha_n}{\!\left\{ \hat{m}, \, \hat{\varpi} \right\} ,\, \alpha_{\backslash n}}
    & \, \propto \,
    \pcond{\alpha_n}{\alpha_{\backslash n}} \!\!\!
    \prod_{i \, \in \, \mathrm{stars\, in\,} n}
    \int
    \tilde{p}_i \arg{\mu_i, \, E \arg{\alpha_n, \mu_i}}
    \, \mathrm{d}\mu_i
    \, ,
    \label{eqn:los-posterior-conditional}
\end{align}
where $n$ is the sightline we are updating, and $\backslash n$ denotes the set of sightlines we are holding fixed. One useful property of a multivariate Gaussian is that the conditional probability densities are also Gaussian. If we fix the sightlines $\backslash n$, the conditional prior on sightline $n$, $\pcond{\alpha_n}{\alpha_{\backslash n}}$, is itself a Gaussian.

To update the central sightline, we hold all the neighboring sightlines fixed, and sample from the conditional distribution on the reddening in the central sightline. We use Markov Chain Monte Carlo sampling to explore the space of reddening densities along the central sightline, holding the neighbors fixed.

To update each neighboring sightline, we hold all the other neighbors and the central sightline fixed. We choose a new sample for the neighboring sightline, from the set of samples stored in the previous iteration. We wish the sample we draw to be distributed according to Eq. \eqref{eqn:los-posterior-conditional}, with $n$ representing the neighbor to be updated, and $\backslash n$ representing all the other sightlines. However, the stored samples of $\alpha_n$ from the previous iteration are drawn from the posterior on $\alpha_n$ in the previous iteration, which we will denote $p_{\mathrm{prev}} \arg{\alpha_n}$. We therefore have to weight each stored sample according to the ratio of the conditional probability density of that sample, Eq. \eqref{eqn:los-posterior-conditional}, to the posterior of that sample from the previous iteration:
\begin{align}
    w \arg{\alpha_n}
    &=
    \frac{
        \pcond{\alpha_n}{\!\left\{ \hat{m}, \, \hat{\varpi} \right\} ,\, \alpha_{\backslash n}}
    }{
        p_{\mathrm{prev}} \arg{\alpha_n}
    }
    \, .
\end{align}
This is essentially an importance-sampling step, with the conditional probability density (Eq. \ref{eqn:los-posterior-conditional}) playing the role of $p \arg{\theta_j}$, and the stored posterior density of the sample from the previous iteration playing the role of $p^{\prime} \arg{\theta_j}$ in Eq. \eqref{eqn:importance-weight}. The details of how we calculate the ratio of these probability densities is given in Appendix \ref{app:neighbor-round-robin}.

The importance-sampling step for the neighboring sightlines works well as long as the samples stored from the previous iteration are a decent representation of the conditional probability density of each neighbor, holding all the other neighbors and the central sightline fixed. If we adiabatically change the correlated prior from iteration to the next, then this is generally the case.

In each iteration, we jointly sample all the sightlines in a small patch surrounding the central sightline, alternating between updating the central sightline and updating each neighbor. We assess convergence by monitoring the autocorrelation of the reddening in the central sightline, as detailed in Appendix \ref{app:convergence}. In order to speed up convergence, we use a technique that is similar to parallel tempering, with an ensemble of samplers that explore modified versions of the joint posterior on reddening density (see Appendix \ref{app:ladder}).

\subsubsection{Implementation details}
\label{sec:implementation-details}

\begin{deluxetable}{c|ccccc}
    \tablecaption{Covariance kernel parameters \label{tab:cov-kernel}}
    \tablehead{
        \colhead{iteration} & \colhead{0} & \colhead{1} & \colhead{2} & \colhead{3} & \colhead{4}
    }
    \startdata
    $\ell$ (pc) & 0 & 0.375 & 0.75 & 1.125 & 1.5 \\
    $r_0$ (pc) & -- & 0.75 & 0.75 & 0.75 & 0.75 \\
    $\ln A$ & -- & -2 & -1 & -\nicefrac{2}{3} & -\nicefrac{1}{2} \\
    \enddata
\end{deluxetable}

In order to impose the Gaussian process prior, we choose the 32 nearest neighbors of each sightline. Each sightline in our map is thus sampled from a slightly different model, in which it is correlated with a patch of sky surrounding it. In each iteration, we sample the patch, as described above, and then discard the information from the neighboring sightlines in the patch, retaining only the Markov chain samples from the central sightline. After an initial pass over the sky, in which we sample the reddening along each sightline independently, we conduct four iterations with a correlated prior, increasing the correlation length with each iteration (see Table \ref{tab:cov-kernel}).

As described above, as we sample the reddening field in a patch of sky, we alternate between updating the central sightline and updating the neighbors. We call one such set of updates a ``round.'' In each round, we make 24000 Metropolis-Hastings proposals in the central sightline (equivalent to 20 proposals per distance bin), and five Gibbs steps in each of the neighboring sightlines. In the initial, uncorrelated iteration, we sample for 2500 rounds, discarding the first 500 as burn-in. In subsequent, correlated iterations, we sample for 2250 rounds, discarding the first 450 as burn-in. We save only the results from the central sightline, storing 100 samples.

The typical runtime per sightline is a function both of the number of stars in the sightline, as well as the number of neighboring sightlines. The runtime contains a term that is linear in the number of stars, due to the need to calculate one line integral per star in the line-of-sight reddening posterior, Eq. \eqref{eqn:los-posterior-final}. For reasons that may have to do with the need to store gridded stellar posterior densities in memory (specifically, with the number that can be stored in cache), there is also a small quadratic dependence of runtime on the number of stars. The runtime has a term that scales quadratically with the number of sightlines, due to the need to sample each neighboring sightline, and that the prior on each sightline depends on all of the neighbors. With 32 neighbors, running on a single core of an Intel Xeon ES-2683v4, with a clock speed of 2.1~GHz and a cache size of 40~MB, we achieve a typical per-sightline runtime of
\begin{align}
    t_{\mathrm{run}} / s
    \approx
    150
    + 0.5 \, n_{\star}
    + 5 \left( \frac{n_{\star}}{100} \right)^{\! 2}
    \, ,
\end{align}
where $n_{\star}$ is the number of stars in the sightline. In the initial iteration, in which no neighboring pixels must be sampled, the typical runtime has the same dependence on the number of stars, but lacks the constant term. This runtime is dominated by the line-of-sight sampler, as the time required for the grid evaluation of the individual stellar posterior densities is negligible, in comparison. The median number of stars per sightline is $\sim$130, yielding a typical runtime of $\sim \! 75 \, \mathrm{s}/\mathrm{sightline}$ for the initial, uncorrelated iteration, and $\sim \! 225 \, \mathrm{s}/\mathrm{sightline}$ for the correlated iterations.

Running on a mix of Intel and AMD cores on Harvard's Odyssey cluster, the initial, uncorrelated iteration consumed approximately 700 million CPU seconds, while each successive iteration consumed approximately 1.5 billion CPU seconds, for a total runtime of $\sim$7 billion CPU seconds to generate the full map.

The method presented here makes a number of approximations to the Gaussian process model. Firstly, it only takes into account correlations between voxels which are at close angular separations on the sky, as seen from the Solar system, thereby giving the Sun a somewhat special position in our priors. Secondly, our voxels are highly elongated along the radial direction, and we calculate the distances between voxels using the distances between their centers. For this reason, distances between voxels are much smaller in the transverse directions than in the radial direction, and the correlations are correspondingly stronger along the transverse directions. We treat the correlations along the radial direction heuristically, through a modification of the inverse covariance matrix, as explained in Appendix \ref{app:ladder}. Nevertheless, this method provides a relatively computationally inexpensive way to approximately apply a Gaussian Process prior to our map, and as will be seen in the following, provides a significantly improved result.

\section{Data}
\label{sec:data}

\subsection{Gaia}
\label{sec:Gaia}

The European Space Agency's \textit{Gaia} spacecraft aims to map the 3D positions and kinematics -- as well as spectral types -- of a substantial fraction of Milky Way stars \citep{GaiaCollaboration2016Mission}. The mission will eventually provide astrometry (positions, proper motions and parallaxes) and optical spectrophotometry for over a billion stars \citep{Liu2012Spectrophotometry}, as well as radial velocity measurements of more than 100 million stars \citep{Katz2004RVS,Wilkinson2005RVS}. Gaia DR2 contains five-parameter astrometric determinations for 1.3 billion sources \citep{Lindegren2018DR2Astrometry}.

Gaia's spectrophotometry is an unusual middle ground between many-band photometry and low-resolution spectroscopy. Gaia takes dispersed images of sources in two bands, RP (330--680~nm) and BP (630--1050~nm). Gaia Data Release 2 \citep{GaiaCollaboration2018DR2Contents} provides integrated RP and BP photometry of 1.4 billion sources is provided, which \citet{Andrae2018DR2Apsis} combines with parallax and $G$-band photometry to estimate effective temperatures for 161 million stars and reddenings for 88 million stars. In future data releases, low-resolution RP and BP spectra will allow better characterization of the parameters of bright stars.

For our work here, the most valuable dataset provided by Gaia DR2 consists of the stellar parallax measurements. Parallaxes provide an independent means of estimating stellar distances from template fitting of spectral energy distributions. Particularly for bright and for nearby stars, Gaia parallaxes can much better constrain stellar distance than photometry alone. Stellar distance-reddening posterior densities based on photometry are often bi-modal, as main-sequence and giant stars of the same temperature can lie at vastly different distances. Even relatively uncertain parallaxes can be sufficient to break this degeneracy (see Fig. 1 of \citealp{Zucker2019MolecularCloudsDR2}).

We make two quality cuts on the Gaia DR2 catalog recommended by \citet{Arenou2018CatalogValidation}:
\begin{align}
    &\bullet \quad \mathtt{visibility\_periods\_used} > 8 \, , \\
    &\bullet \quad \frac{\mathtt{astrometric\_chi2\_al}}{\mathtt{astrometric\_n\_good\_obs\_al} - 5} < \chi_{\mathrm{threshold}}^2 \, ,
\end{align}
where $\chi_{\mathrm{threshold}}^2$ is related to the Gaia $G$-band magnitude, $m_G$, by
\begin{align}
    \chi_{\mathrm{threshold}}^2
    &=
    1.44 \, \max \left\{
        1, \ 
        \exp \left[
            -0.4 \left( m_{G} - 19.5 \right)
        \right]
    \right\} \, .
\end{align}
The first of these cuts removes sources with an insufficient number of Gaia observations to safely make a parallax measurement, while the second cut removes sources which are poorly fit by the astrometric model (and is essentially a cut on reduced $\chi^2$).

\subsection{Pan-STARRS 1}
\label{sec:PS1}

The bulk of our stellar photometry comes from the Panoramic Survey Telescope and Rapid Response System 1 (Pan-STARRS~1, hereafter abbreviated as ``PS1''), a 1.8~m optical and near-infrared telescope located on Mount Haleakala, Hawaii \citep{Chambers2016PS1}. The telescope is equipped with the Gigapixel Camera \#1 (GPC1), consisting of an array of 60 CCD detectors, each 4800 pixels on a side \citep{Tonry2006-GPC1,Onaka2008-GPC1,Chambers2016PS1}. From May 2010 to April 2014, the majority of the observing time was dedicated to a multi-epoch survey of the sky north of declination $\delta = -30^{\circ}$ \citep[named the ``$3 \pi$ survey'' for its footprint in steradians]{Chambers2016PS1}. The $3 \pi$ survey observed in four SDSS-like passbands \citep{York2000}, $\gps$, $\rps$, $\ips$ and $\zps$, as well as an additional passband in the near-infrared, $\yps$. The entire filter set spans the range 400--1000~nm \citep{Stubbs2010-PS1-lasercal}.

Astrometry and photometry was extracted by the PS1 Image Processing Pipeline \citep{Magnier2016-PS1-data-processing,Magnier2016-PS1-pixel-analysis,Magnier2016-PS1-photo-astro-calib}. PS1 photometry features a very uniform flux calibration, achieving better than 1\% precision over the sky \citep{Schlafly2012,Chambers2016PS1}. In single-epoch photometry, the $3 \pi$ survey reaches typical $5\,\sigma$ depths of 22.0~mag (AB) in $\gps$, 21.8~mag in $\rps$, 21.5~mag in $\ips$, 20.9~mag in $\zps$, and 19.7~mag in $\yps$ \citep{Chambers2016PS1}. The PS1 Data Release 1 (DR1) occurred in December 2016, and provided a static-sky catalog, stacked images from the $3 \pi$ survey, along with other data products \citep{Flewelling2016-PS1-database}.

Because of its large footprint, homogeneous depth and excellent internal calibration, PS1 photometry provides an ideal dataset for mapping Galactic dust through stellar colors. As in \citet{Green2018}, we use stellar photometry derived from the single-epoch photometry, which is very similar to PS1 DR1, with the primary difference being in the treatment of observations taken in non-photometric conditions.

\subsection{Two-Micron All-Sky Survey} \label{sec:2MASS}

The Two Micron All Sky Survey (2MASS) is an all-sky survey in three near-infrared passbands, $J$, $H$ and $K_{s}$ \citep{Skrutskie2006}, conducted from two 1.3~m telescopes at Mount Hopkins, Arizona and Cerro Tololo, Chile. The name of the survey derives from the wavelength range covered by the reddest passband, $K_{s}$. Ground-based surveys beyond the atmospheric window at $2 \, \mathrm{\mu m}$ are hampered by severe background thermal emission \citep{Skrutskie2006}. The focal plane of each telescope was equipped with three $256 \times 256$ pixel arrays, with a $2^{\prime}$ pixel scale. The entire sky was covered six times with dual 51-millisecond and 1.3-second exposures, in order to cover a wide range in apparent magnitudes. The survey achieved a 10-$\sigma$ point-source depth of $\sim$15.8, $\sim$15.1 and $\sim$14.3~mag (Vega) in $J$, $H$ and $K_{s}$, respectively. 2MASS photometry was calibrated to 0.02~mag accuracy, with per-source photometric uncertainties for bright sources below 0.03~mag \citep{Skrutskie2006}.

In this work, as in \citet{Green2018}, we make use of the ``high-reliability catalog,''\footnote{See the \textit{Explanatory Supplement to the 2MASS All Sky Data Release}: \url{https://old.ipac.caltech.edu/2mass/releases/allsky/doc/sec1\_6b.html\#composite}} and exclude sources that are marked as possibly having contamination from nearby point sources or galaxies.

\subsection{Input Catalog}
\label{sec:input-catalog}

\begin{deluxetable}{ccccc}
    \tablecaption{Input catalog source statistics \label{tab:source-stats}}
    \tablehead{
        \multicolumn{2}{c}{\% sources detected} & & \multicolumn{2}{c}{\% sources detected in} \\
        \multicolumn{2}{c}{in each band} & \quad & \multicolumn{2}{c}{$N$ photometric bands} \\ \cline{1-2} \cline{4-5}
        \colhead{band} & \colhead{\%} & & \colhead{$N$} & \colhead{\%}
    }
    \startdata
    $\gps$           & 79.0 & & 4  & 29.1 \\
    $\rps$           & 98.1 & & 5  & 52.5 \\
    $\ips$           & 97.0 & & 6  & 4.7  \\
    $\zps$           & 98.0 & & 7  & 5.0  \\
    $\yps$           & 93.8 & & 8  & 8.7  \\
    $J$              & 18.5 & & -- & --     \\
    $H$              & 15.5 & & -- & --     \\
    $K_{\mathrm{S}}$ & 11.6 & & -- & --     \\
    $\varpi$         & 62.4 & & -- & -- \\
    $\varpi / \sigma_{\varpi} > 5$ & 8.6 & & -- & -- \\
    $\varpi + 2 \sigma_{\varpi} < 1 \, \mathrm{mas}$ & 32.4 & & -- & -- \\[2pt]
    \tableline
    \multicolumn{5}{l}{799 million sources total.} \\
    \enddata
\end{deluxetable}

\begin{deluxetable}{ccccc}
    \tablecaption{Pixelization of the sky \label{tab:sky-pixelization}}
    \tablehead{
        \multirow{2}{*}{\texttt{nside}} &
        pixel &
        Max. &
        $\Omega$ &
        \# of \\
        &
        scale &
        stars/pix &
        ($\mathrm{deg}^{2}$) &
        pixels
    }
    \startdata
    64    &  $55^{\prime}$   &  40   &  49     &  58       \\
    128   &  $27^{\prime}$   &  40   &  63     &  302      \\
    256   &  $14^{\prime}$   &  40   &  249    &  4756     \\
    512   &  $6.9^{\prime}$  &  600  &  23019  &  1755295  \\
    1024  &  $3.4^{\prime}$  &  ---  &  8044   &  2453659  \\[2pt]
    \tableline
    totals &  ---  &  ---  &  31425  &  4214070  \\[2pt]
    \enddata
\end{deluxetable}

We generate a matched input catalog combining PS1 $grizy$ photometry, 2MASS $JHK_{s}$ photometry and Gaia parallaxes, using the \textit{Large Survey Database} framework \citep{Juric2012LSD}. We use a matching radius of $1^{\prime \prime}$ to match 2MASS sources to PS1, and a radius of $0.2^{\prime \prime}$ to match Gaia DR2 sources to PS1. We require that each source be observed in at least four photometric passbands, at least two of which must be from PS1. We do not require that sources have measured parallaxes. To exclude extended sources, we require that the ${\hat{m}_{\mathrm{PSF}} - \hat{m}_{\mathrm{ap}} < 0.1}$ in at least two PS1 bands, where $\hat{m}_{\mathrm{PSF}}$ are magnitudes determined using PSF photometry, and $\hat{m}_{\mathrm{ap}}$ are aperture magnitudes. Our resulting catalog contains 799 million sources, of which 19.5\% are observed in at least one 2MASS passband, 62.4\% have a Gaia parallax measurement, and 8.6\% have a Gaia parallax measurement with ${\nicefrac{\hat{\varpi}}{\sigma_{\varpi}} > 5}$. It should be noted, however, that even parallax measurements that do not confidently rule out zero parallax can put useful lower limits on distance. In our input catalog, 32.4\% of sources are constrained at ${> \! 2 \sigma}$ by their Gaia parallaxes to be more distant than 1~kpc.

We divide up the sky using a multi-resolution HEALPix pixelization \citep{Gorski2005}. We assign the sources to angular pixels in the same manner as in \citet{Green2018}, beginning with a coarse angular pixelization of the sky and recursively subdividing pixels with more than a given number of sources. Given that our spatially correlated priors allow information to be shared between nearby voxels, we more aggressively subdivide pixels than in \citet{Green2018}, as reflected in the subdivision thresholds in Table \ref{tab:sky-pixelization} (``max. stars / pixel''), resulting in 4.21 million pixels (versus 3.42 million over the same footprint in our previous work). The properties of our input catalog are summarized in Tables \ref{tab:source-stats} and \ref{tab:sky-pixelization}.

\section{Results} \label{sec:results}

\begin{figure*}
    \centering
    \includegraphics[width=\textwidth]{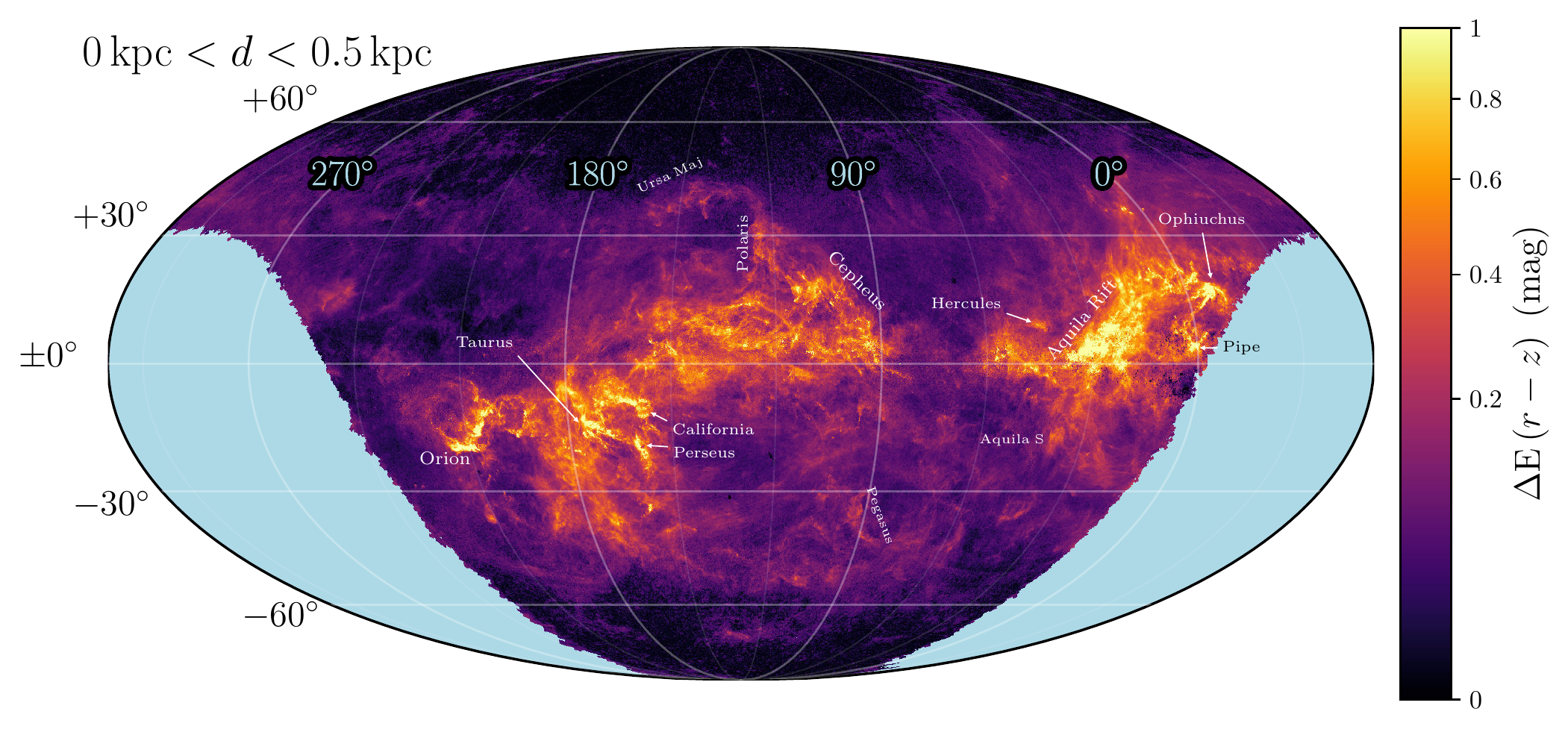}
    \caption{Cumulative reddening out to 500~pc.}
    \label{fig:allsky-diff-0}
\end{figure*}

\begin{figure*}
    \centering
    \includegraphics[width=\textwidth]{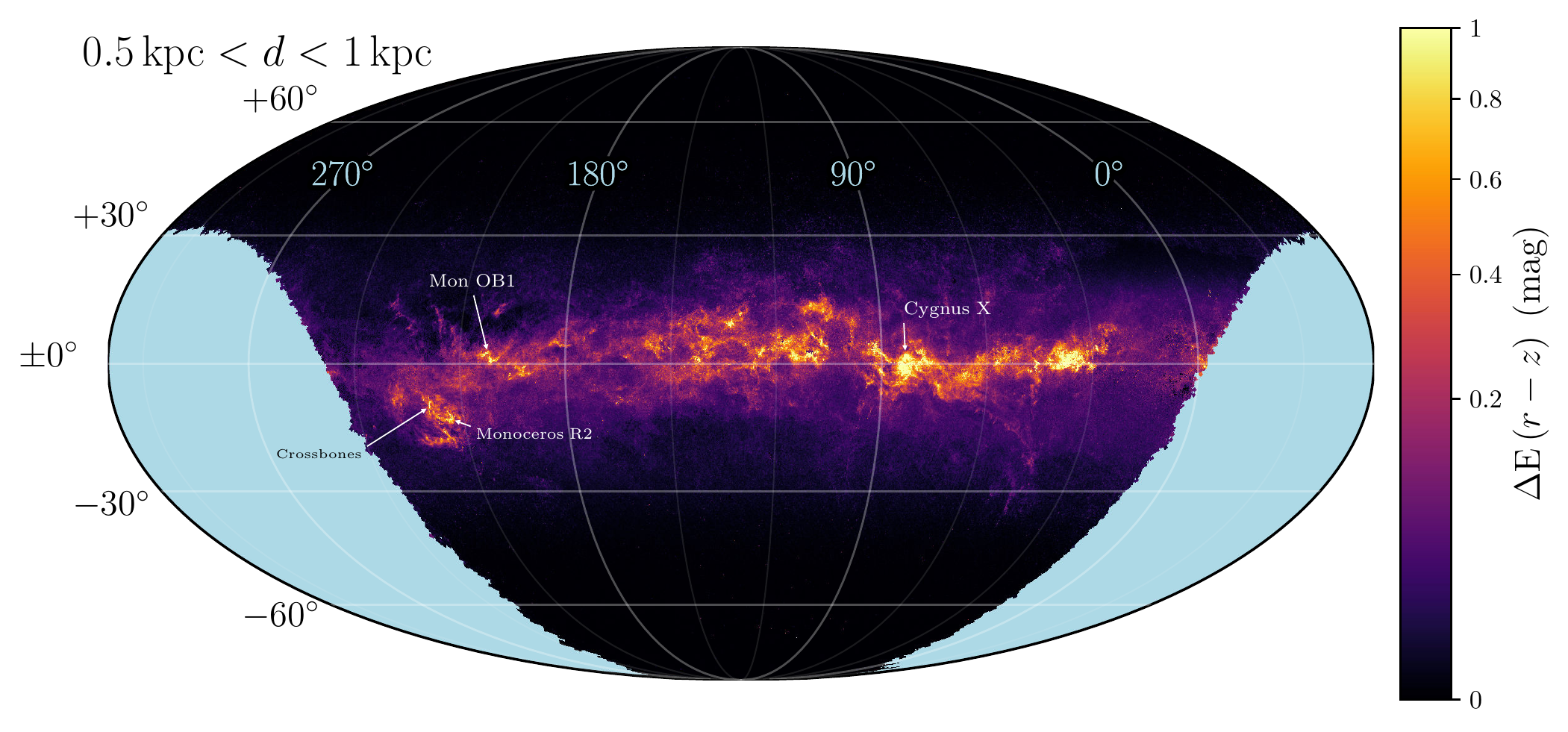}
    \caption{Cumulative reddening between 0.5 and 1~kpc.}
    \label{fig:allsky-diff-1}
\end{figure*}

\begin{figure*}
    \centering
    \includegraphics[width=\textwidth]{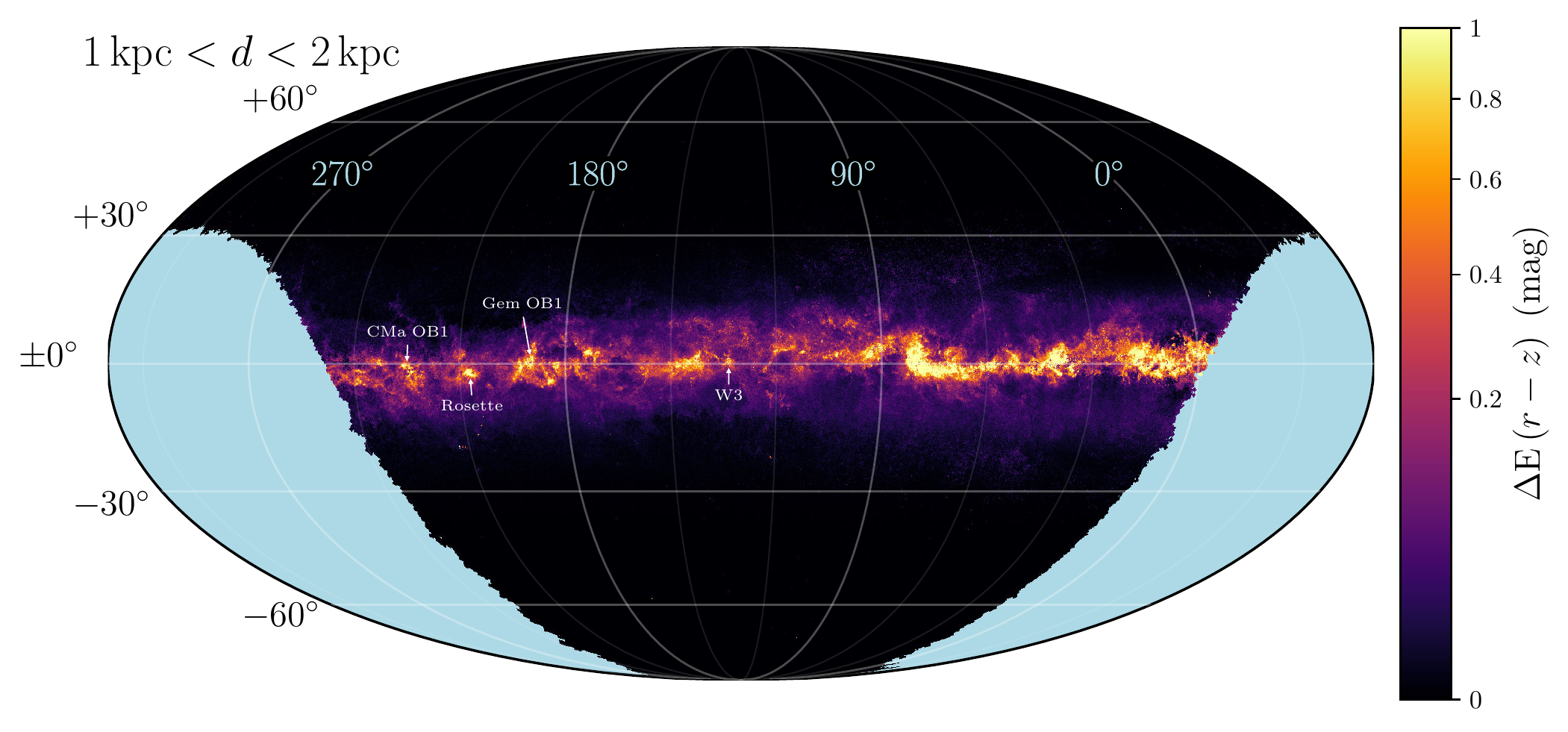}
    \caption{Cumulative reddening between 1 and 2~kpc.}
    \label{fig:allsky-diff-2}
\end{figure*}

\begin{figure*}
    \centering
    \includegraphics[width=\textwidth]{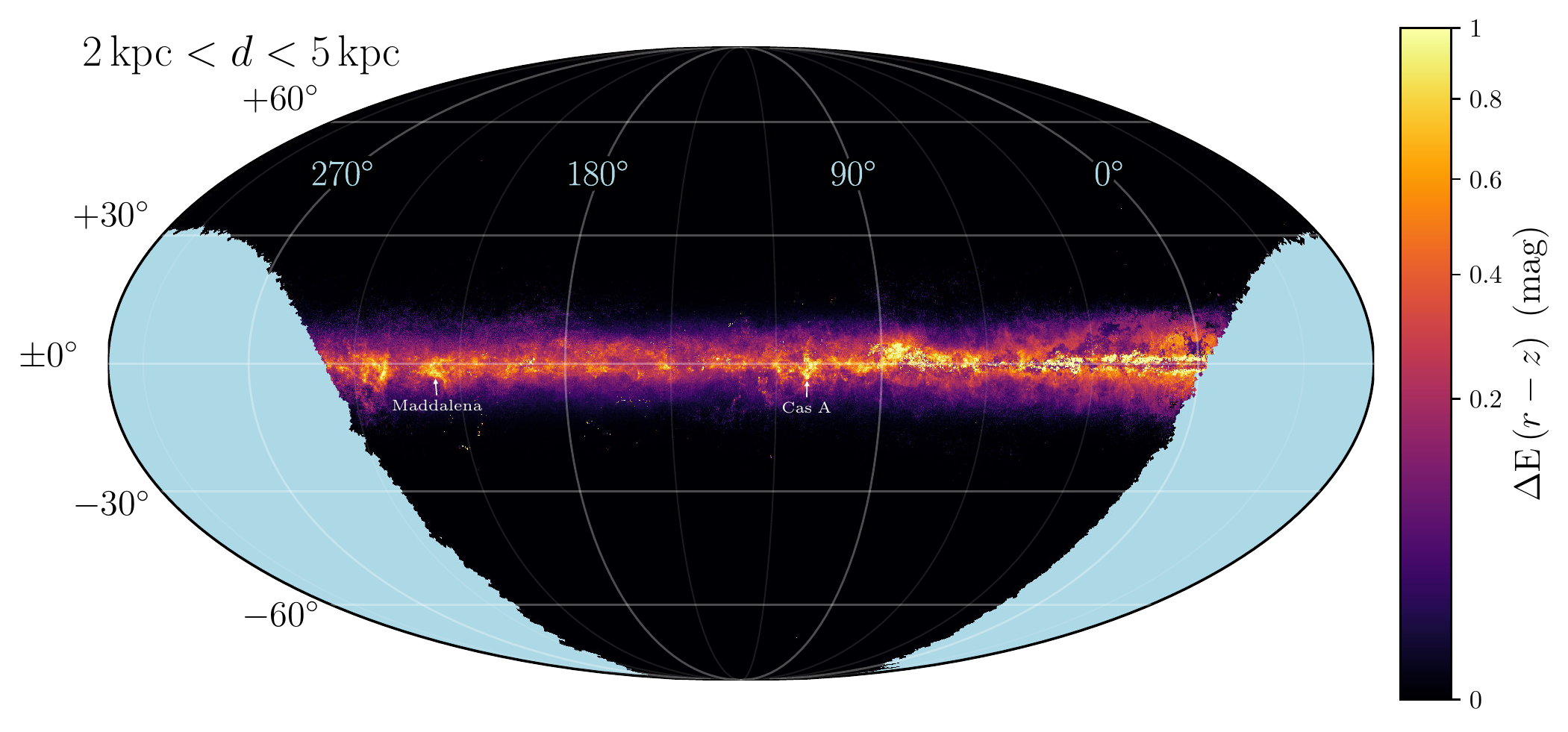}
    \caption{Cumulative reddening between 2 and 5~kpc.}
    \label{fig:allsky-diff-3}
\end{figure*}

\begin{figure*}
    \centering
    \includegraphics[width=\textwidth]{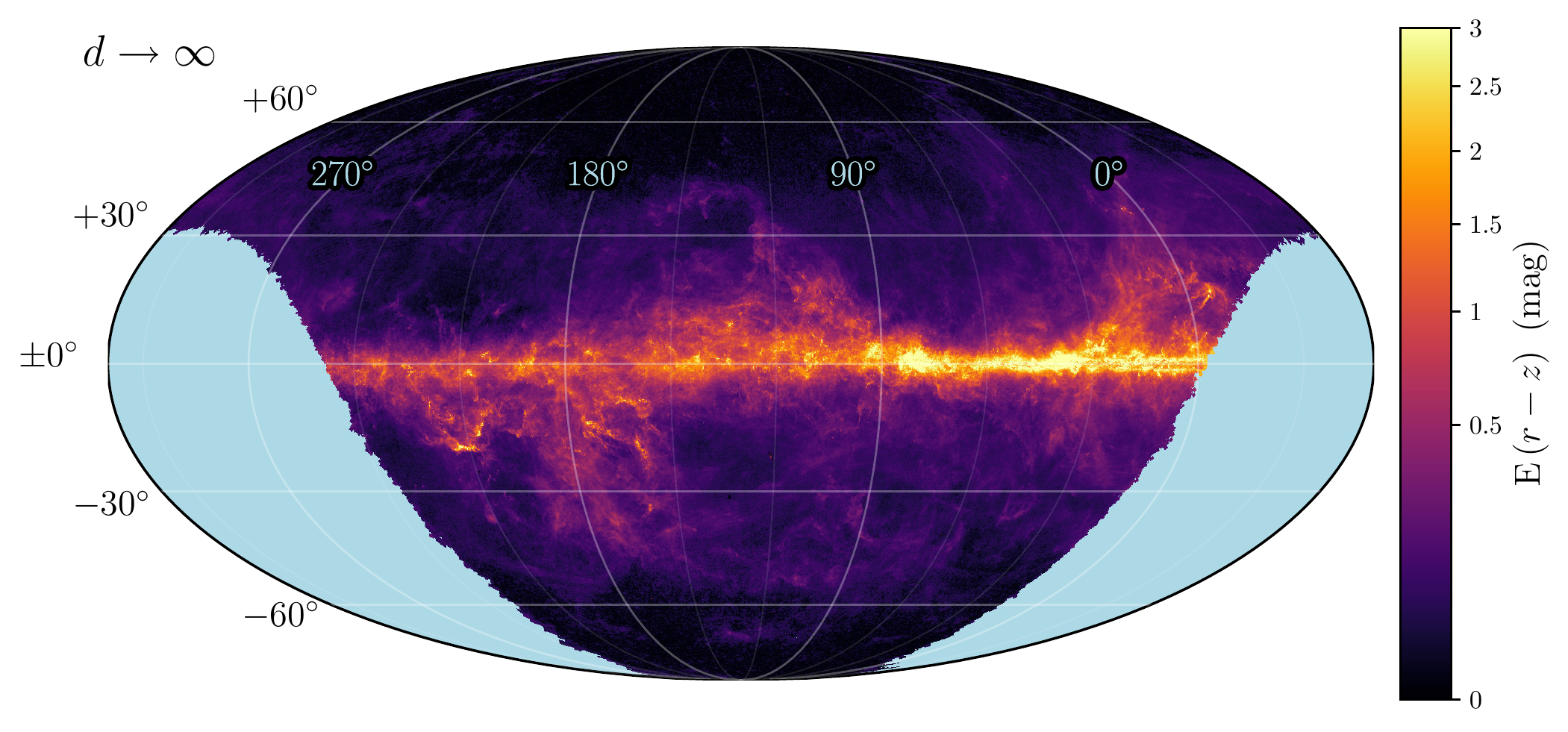}
    \caption{Total integrated reddening.}
    \label{fig:allsky-cumulative}
\end{figure*}

We obtain a map of dust reddening density, with angular sightlines of a typical scale of $3.4^{\prime}$ to $13.7^{\prime}$, and 120 distance bins spaced logarithmically in distance from 63~pc to 63~kpc. Figs. \ref{fig:allsky-diff-0}--\ref{fig:allsky-diff-3} show differential reddening in four distance ranges, out to 5~kpc, while Fig. \ref{fig:allsky-cumulative} shows the cumulative reddening out to the maximum distance in the map. Our map is probabilistic, and we obtain samples of the reddening density in each voxel. In order to transform this into a single value for plotting purposes, we first construct a map of median integrated reddening out to each voxel. When we wish to display reddening densities, we use the line-of-sight derivative of this ``median'' map.

Figs. \ref{fig:allsky-diff-0}--\ref{fig:allsky-diff-3} reveal a wealth of structure at different scales and at different distances. The nearest dust structures appear off of the Galactic plane in Fig. \ref{fig:allsky-diff-0} -- Orion, Taurus, Perseus, California, the Aquila Rift and $\rho$ Ophiuchus. In Fig. \ref{fig:allsky-diff-1}, Monoceros R2 and Cygnus X are clearly visible. At increasing distances, in Figs. \ref{fig:allsky-diff-2}--\ref{fig:allsky-diff-3}, dust features are increasingly confined to low Galactic latitude, as should be expected for the Galactic disk.

\subsection{Dependence on correlation length}

\begin{figure*}
    \centering
    \includegraphics[width=\textwidth]{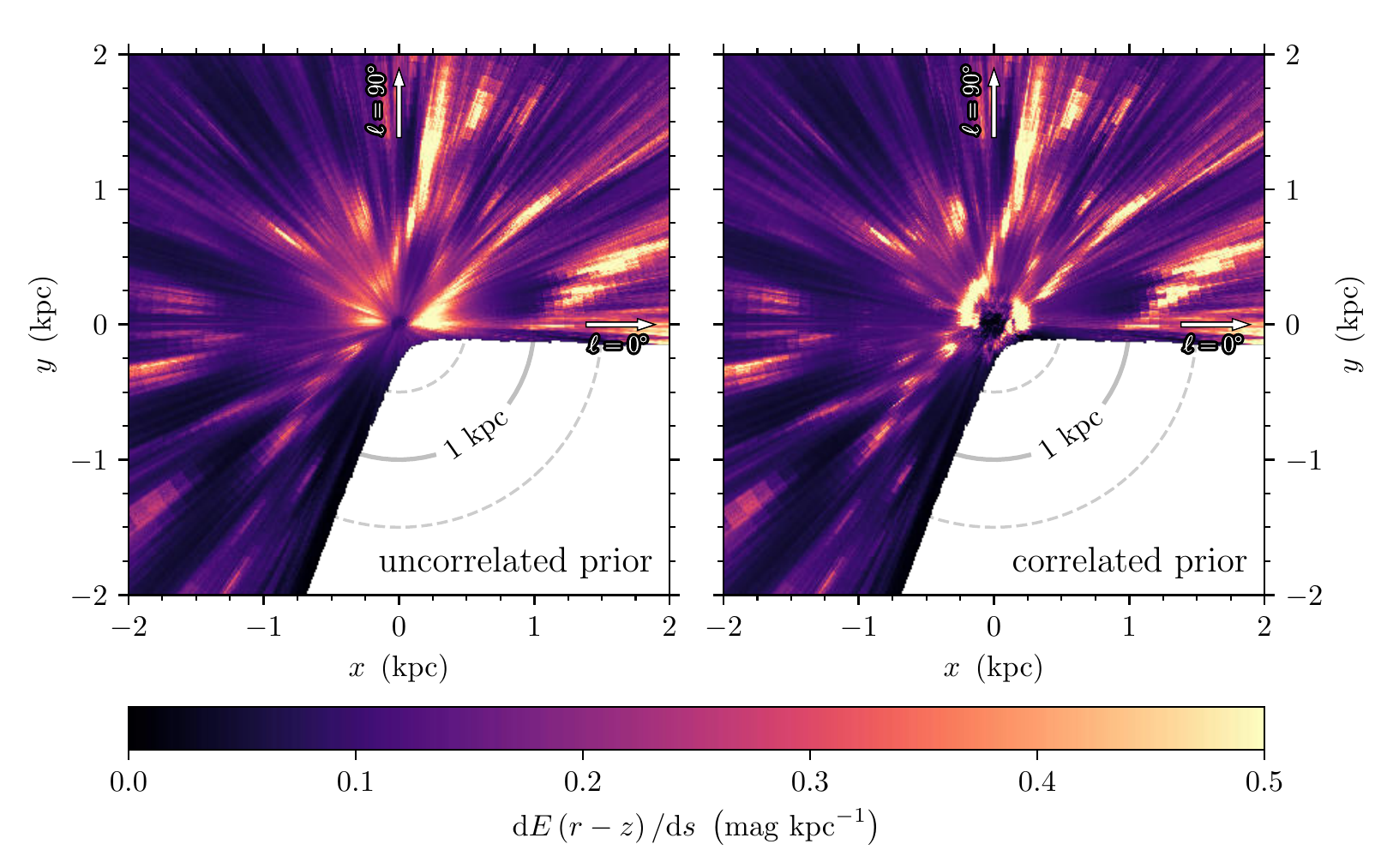}
    \caption{A face-on view of the Galaxy, with the Sun at the center. The right of each panel corresponds to $b=0^{\circ}$, and the Galactic center lies off the right edge of the plot. The left panel shows the initial iteration, without a spatially correlated prior on dust reddening density, while the right panel shows the final iteration, with a correlation length of 1.5~pc. We integrate the reddening from $z = -300 \, \mathrm{pc}$ to $+300 \, \mathrm{pc}$. As is evident from a comparison of the two panels, the correlated prior yields more isotropic cloud shapes, with less smearing along the line of sight. This effect is especially pronounced at nearby distances, where neighboring sightlines are more tightly coupled.}
    \label{fig:birdseye-iterations}
\end{figure*}

\begin{figure*}
    \centering
    \includegraphics[width=0.98\textwidth]{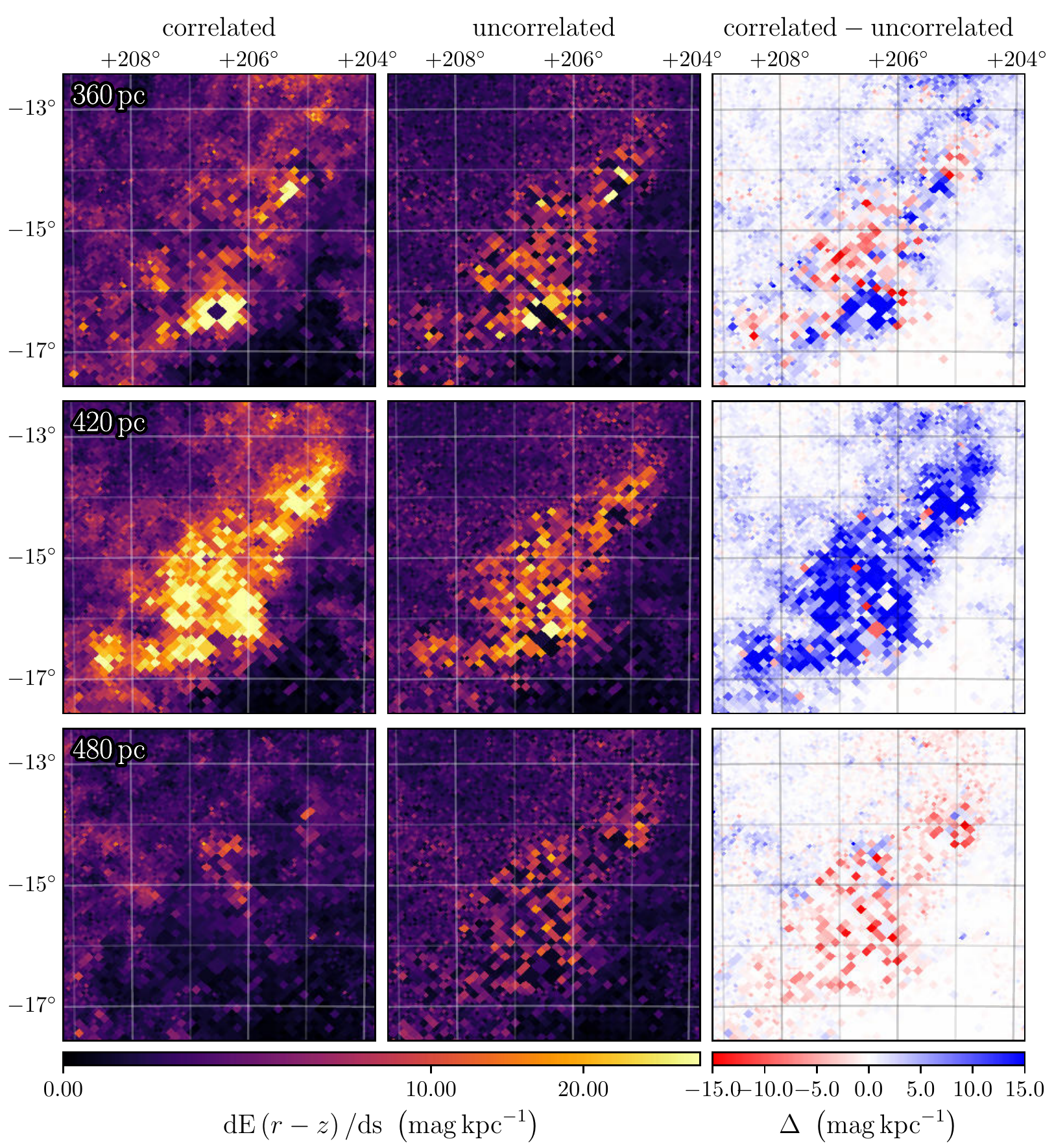}
    \caption{Reddening density in the vicinity of Orion B. Each row corresponds to a different distance. The left column shows the results from the final iteration, with a correlation length of 1.5~pc. The center column shows the results from the initial, uncorrelated iteration, while the right column shows the difference between the final and inital iterations. Dust clouds have fewer ``holes'' when the correlated prior is applied, and are better localized in distance. Note that in the correlated iteration, we localize the distance of Orion B much more precisely, while in the uncorrelated iteration, it is smeared across a greater range of distances.}
    \label{fig:orion-b-zoomin}
\end{figure*}

The Gaussian process prior, which couples the reddening density in nearby voxels, has a significant impact on the recovered reddening map. We can see this by comparing the results from the first iteration, in which the sightlines are independent, and the final iteration, in which the sightlines are coupled, as shown in Fig. \ref{fig:birdseye-iterations}. The correlated prior favors dust clouds which are isotropic, rather than stretched along the line of sight. As the transverse distance between neighboring sightlines grows with distance, the correlations between them are stronger in the nearest distance bins than in the farthest distance bins. The effect of our Gaussian process prior is thus greatest for the nearest clouds, particularly within a distance of $\sim$1~kpc.

The effect of the Gaussian process prior can be seen in a striking way in close-up views of individual clouds. Fig. \ref{fig:orion-b-zoomin} shows the reddening density in our correlated and uncorrelated maps (corresponding to the final and initial iterations, respectively) in the vicinity of the Orion B molecular cloud complex. The distance to the cloud complex is much better localized in the correlated map, whereas in the uncorrelated map, the dust is spread over a wider range of distances. Slices through the dust density field in the uncorrelated map show holes, caused by the model placing the cloud complex at different distances in neighboring sightlines. This noise is much reduced in the correlated map. The results for Orion B are typical of our results for clouds within ${\lesssim 1}$~kpc.

\subsection{Comparison with 2D dust map: Planck Collaboration (2014)}

Although we infer dust reddening in 3D voxels, we can compare our results with those of 2D dust maps by integrating out the distance direction in our map. In the following, we use the median of our integrated reddening at each location on the sky.

\begin{figure*}
    \centering
    \plottwo{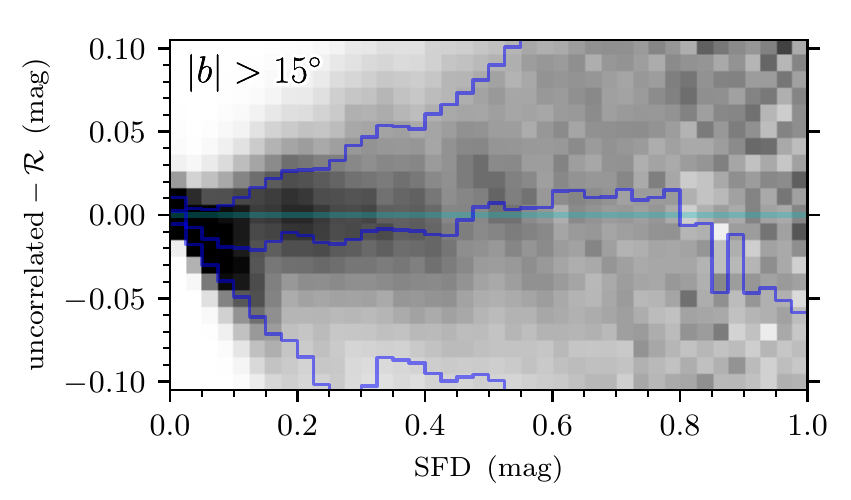}{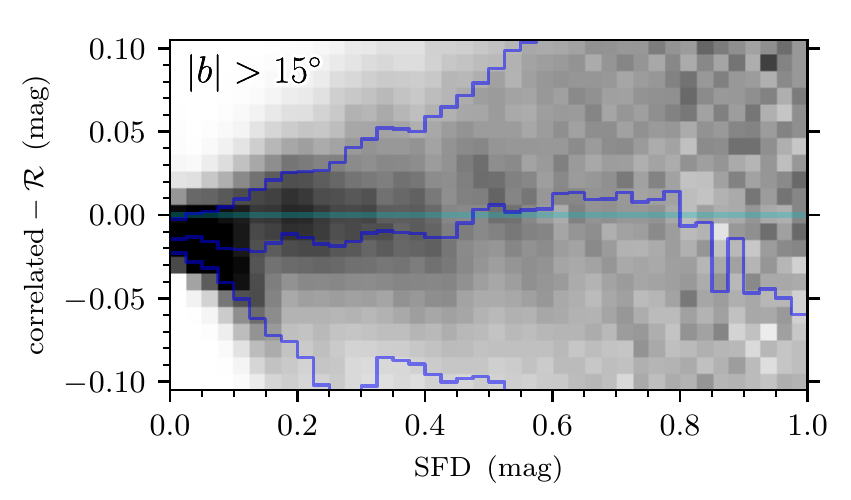}
    \caption{Histogram of the residuals of our map with the Planck14 radiance-based reddening estimate, as a function of the SFD reddening. We exclude the Galactic midplane (defined as $\left| b \right| < 15^{\circ}$). The left panel compares our uncorrelated map (initial iteration) with Planck, while the right panel compares our correlated map (final iteration) with Planck. All reddenings are in magnitudes of $E \arg{g - r}$. The blue lines give the $16^{\mathrm{th}}$ percentile, median and $84^{\mathrm{th}}$ percentile of the reddening residuals at each distance.}
    \label{fig:corr-planck-radiance}
\end{figure*}

\citet[][hereafter ``Planck14'']{PlanckCollaboration2013} fits a modified black-body model of dust emission to the Planck 857, 545 and 353 GHz bands, as well as the IRAS $100 \, \mathrm{\mu m}$ band. Planck14 then constructs two different maps of dust reddening by separately calibrating optical depth at 353 GHz and integrated radiance against reddening measurements of SDSS quasars. While far-infrared optical depth should be a better tracer of dust reddening than integrated radiance is, at high Galactic latitude, optical depth estimates can be noisy, due to covariances between the dust optical depth, temperature and $\beta$ (which describes the shape of the modified black-body spectrum). In these comparisons, We therefore use the reddenings derived from integrated radiance, which we denote as $\mathcal{R}$. We transform these reddenings from $E \arg{B - V}$ to $E \arg{g - r}$ using the ${R_V = 3.1}$ reddening curve of \citet{SchlaflyFinkbeiner2011}.

In Fig. \ref{fig:corr-planck-radiance}, we compare our map to the Planck radiance-based dust map out of the Galactic midplane (${\left| b \right| > 15^{\circ}}$). In the midplane of the Galaxy, we would expect Planck to detect dust out to greater extinctions, as dust remains optically thin in far-infrared emission. By contrast, our method relies on observing stars behind the dust, and therefore cannot trace reddening to as great a depth. We compare the Planck14 map with both the initial iteration of our dust map, without spatially correlated priors, and the final iteration of our dust map, which has spatially correlated priors. For both versions of the map, we find good agreement with Planck14 out to a reddening of ${E \arg{g-r} \approx 1 \, \mathrm{mag}}$, with median residuals between the maps of $\lesssim 0.05 \, \mathrm{mag}$.

At reddenings of $E \arg{g-r} \lesssim 0.1 \, \mathrm{mag}$, the final iteration of our map (with correlated spatial priors) matches Planck14 better than the initial iteration does (with completely independent sightlines), in that the residuals have a flat trend with increasing reddening. At these low reddenings, the trends in the residuals between the two maps are driven by possible zero-point offsets in both maps, as well as the details of the priors we impose on the dust reddening. The introduction of the correlated spatial prior on dust reddening density should act to increase the signal-to-noise ratio in our reddening measurements, as the reddening in each sightline is influenced by stars in nearby sightlines. If our prior favors too much reddening at high Galactic latitudes, then increased signal-to-noise should drive our reddening measurements down, and closer to the correct answer.

\subsection{Comparison with 3D dust maps}

\subsubsection{Green \textit{et al.} (2018)}

\begin{figure*}
    \centering
    \plottwo{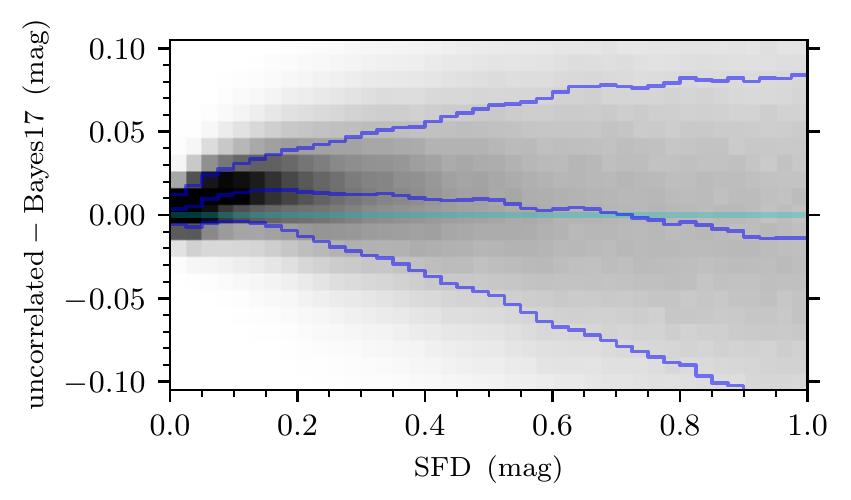}{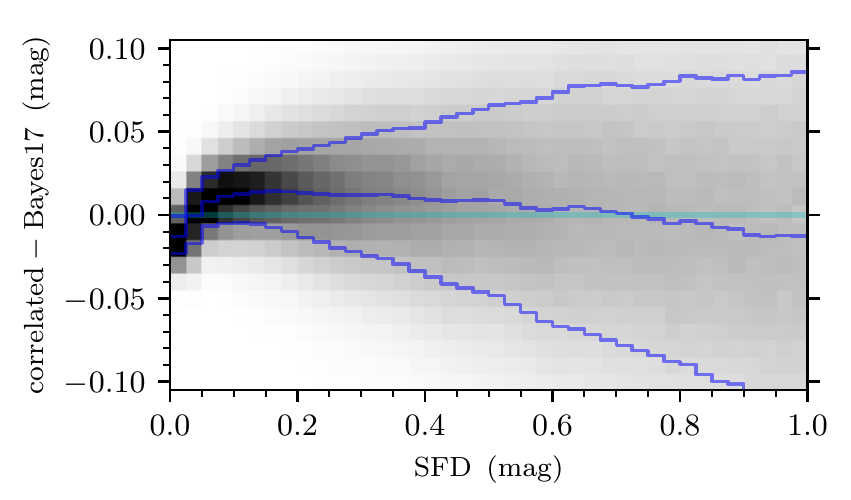}
    \caption{Histogram of the residuals of our new map with Bayestar17, as a function of the SFD reddening. Both maps are integrated to infinite distance. The left panel compares our uncorrelated map (initial iteration) with Bayestar17, while the right panel compares our correlated map (final iteration) with Bayestar17. All reddenings are in magnitudes of $E \arg{g - r}$.}
    \label{fig:corr-bayestar2017}
\end{figure*}

In Fig. \ref{fig:corr-bayestar2017}, we compare our integrated reddenings at large distance (for both the initial and final iterations of our map) with those of \citet[][hereafter ``Bayestar17'']{Green2018}. At reddenings greater than $E \arg{g-r} \sim 0.1 \, \mathrm{mag}$, our results are largely consistent, with median residuals of $\lesssim 0.02 \, \mathrm{mag}$ out to a reddening of $E \arg{g-r} \approx 1 \, \mathrm{mag}$. The major differences between the initial iteration and Bayes17 are the use of Gaia parallaxes in the former, as well as slightly different reddening vectors (particularly in the 2MASS passbands) and additional distance bins in the former, which has the effect of slightly altering our priors. In the final iteration of our dust map, regions of the sky with low reddening differ most from Bayes17 (and from the initial iteration), with more of the sky being placed at the lowest reddenings. The spatially correlated priors effectively cause information to be shared between nearby voxels. If the stellar photometry in one voxel constrains the reddening to be close to zero, this information propagates to the neighboring voxels, pulling down the inferred reddening. If the priors on individual voxels at high Galactic latitude favor more dust than is typically found in these regions, then the introduction of spatially correlated priors will tend to pull these regions closer to the true reddening.

\subsubsection{Chen \textit{et al.} (2019)}

\begin{figure*}
    \centering
    \includegraphics[width=\textwidth]{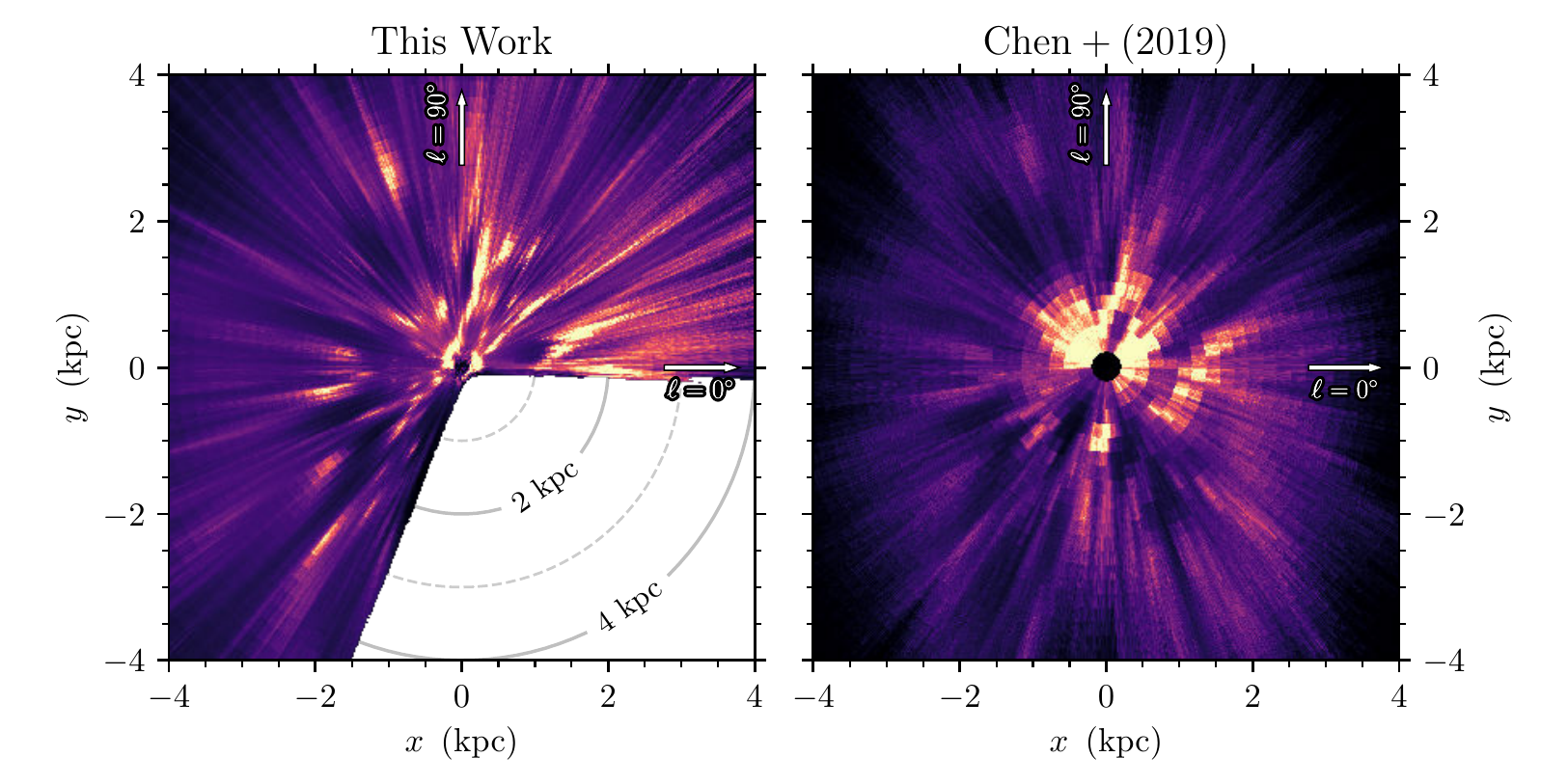}
    \caption{Bird's-eye comparison with \citet{Chen2018}. The Sun is located in the middle of each panel, with the Galactic center off the edge of the plot to the right. Reddening density is integrated between $z = \pm 300 \, \mathrm{pc}$, and is displayed in arbitrary units.}
    \label{fig:birdseye-vs-chen2018}
\end{figure*}

\citet[][hereafter ``Chen19'']{Chen2018} uses stellar parallax measurements from Gaia DR2, as well as optical and near-infrared photometry from Gaia, WISE and 2MASS to trace dust reddening in the Galactic plane ($\left| b \right| < 10^{\circ}$). In Fig. \ref{fig:birdseye-vs-chen2018}, we compare our map with that of Chen19. Due to our use of deeper Pan-STARRS 1 photometry, we trace dust density to greater distances. The greater distance resolution of our map also reveals much finer features. This greater distance resolution is enabled in part by our larger input catalog (799 million versus 56 million stars). The gross features revealed by both maps are similar, increasing our confidence that both maps are recovering real features in the interstellar medium. Neither map shows clear spiral structure -- we discuss the possible relation of the features in our map to spiral structure in Section \ref{sec:spirals}.

\subsubsection{Leike \& En{\ss}lin (2019)}
\label{sec:le2019}

\begin{figure*}
    \centering
    \includegraphics[width=\textwidth]{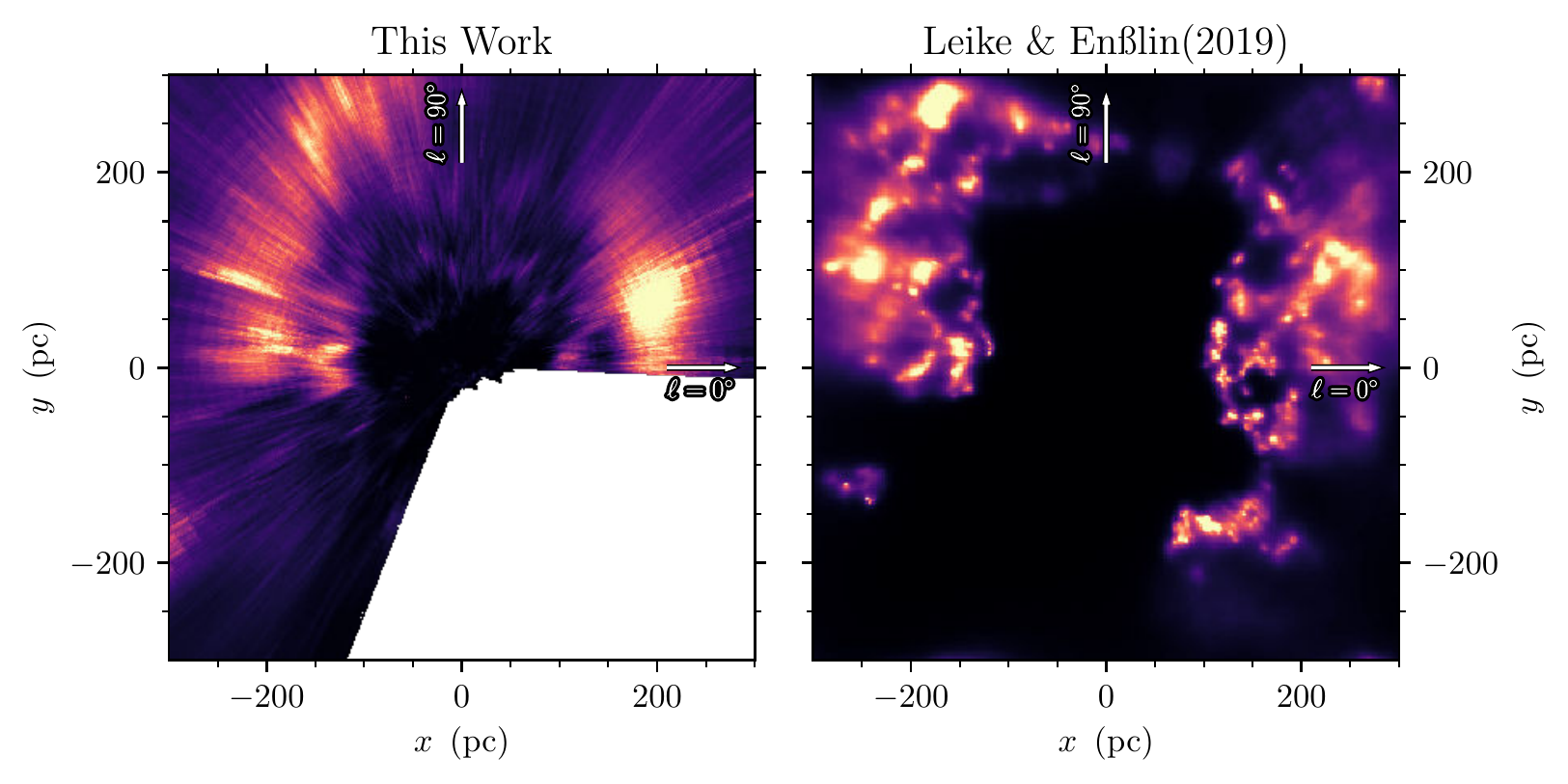}
    \includegraphics[width=\textwidth]{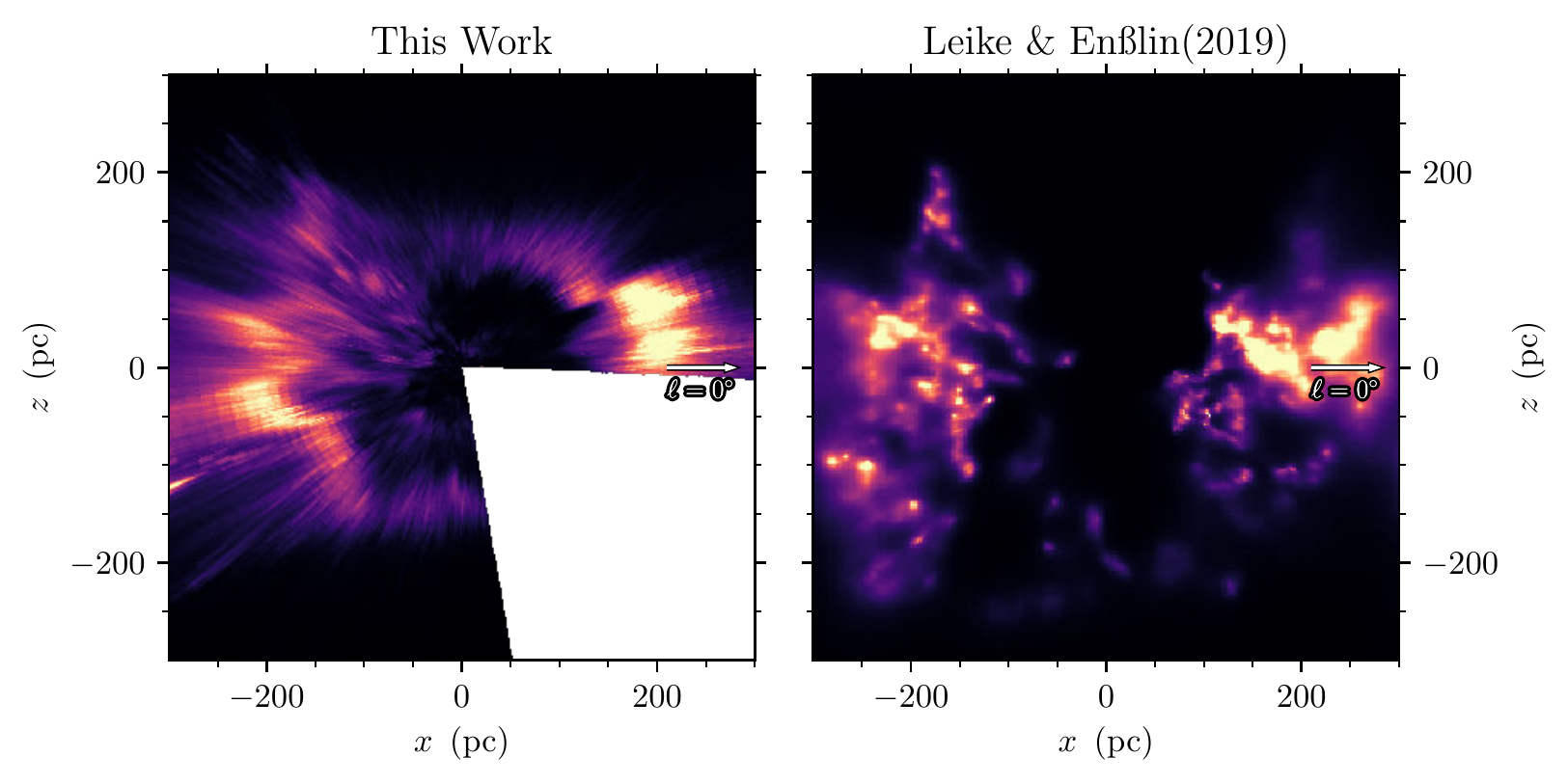}
    \caption{Comparison with \citet{LeikeEnsslin2019}. Reddening density is integrated between $z = \pm 300 \, \mathrm{pc}$ in the top panels, and between $y = \pm 300 \, \mathrm{pc}$ in the bottom panels, and is displayed in arbitrary units. The masked regions in the left panels correspond to regions in which less than 50\% of the volume along the projection axis is filled, due to our footprint not covering declinations below $-30^{\circ}$.}
    \label{fig:birdseye-vs-leikeensslin2019}
\end{figure*}

\begin{figure*}
    \centering
    \includegraphics[width=\textwidth]{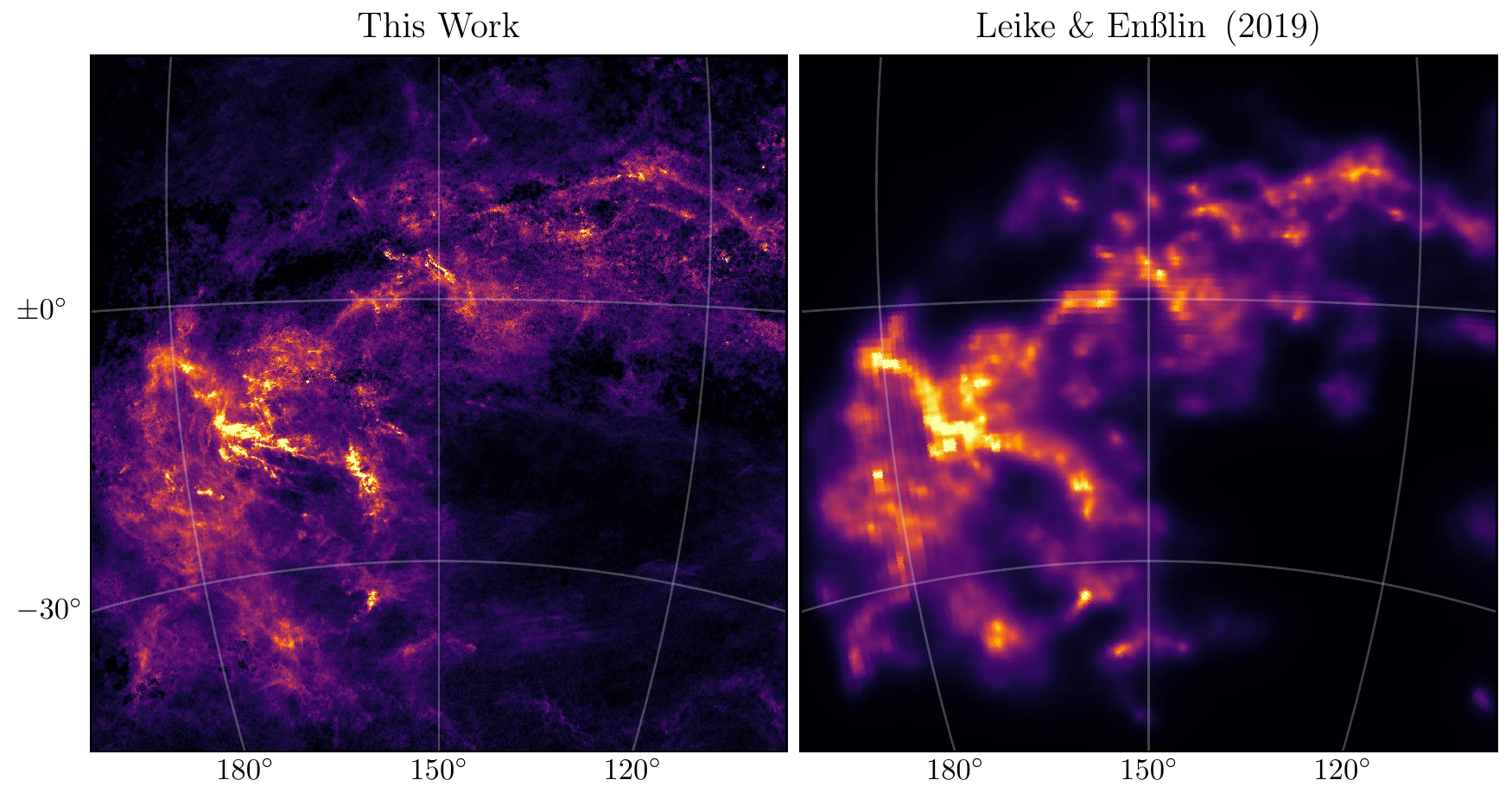}
    \caption{Comparison with \citet{LeikeEnsslin2019} in the Galactic anticenter. Reddening density is integrated out to a distance of $300 \, \mathrm{pc}$, and displayed in arbitrary units.}
    \label{fig:anticenter-vs-leikeensslin2019}
\end{figure*}

\citet[][hereafter ``LE2019'']{LeikeEnsslin2019} reconstructs the three-dimensional structure of nearby dust using Gaia DR2 parallaxes and reddening estimates (the latter from \citealp{Andrae2018DR2Apsis}), treating the logarithm of the dust density as a Gaussian process. LE2019 applies their model to 3.7 million stars with well-determined Gaia parallaxes in a $\left(600 \, \mathrm{pc} \right)^3$-cube centered on the Sun, obtaining the three-dimensional distribution of dust, as well as its spatial correlation spectrum.

There are a number of important differences between our present work and that of LE2019. Unlike in our work, LE2019 uses a hierarchical Bayesian model in which the kernel parameters are inferred, rather than fixed beforehand. Their resulting correlation kernel has correlations on much larger scales than the kernel we assume, on the order of 30~pc. LE2019 uses a constant mean in their Gaussian process prior over all of their volume. The mean of our prior, by contrast, is set to follow a smooth Galactic disk, encoding our prior expectations about the overall structure of the Galaxy. A key difference between our methods is that in the present work, we approximate the Gaussian process prior in small angular patches surrounding each sightline, disregarding correlations between voxels that are separated by more than $\sim \! 22^{\prime}$ in angle on the sky. At a distance of 300~pc, this means that our prior ties voxels that are separated by a maximum physical distance of $\sim \! 1.9 \, \mathrm{pc}$. This produces more spherical clouds than one would get with an uncorrelated prior (see Fig. \ref{fig:birdseye-iterations}), but does not fully exploit the information that is available in spatial correlations between nearby voxels. One consequence of this treatment is that some of the nearest clouds, such as the $\rho$~Ophiuchi and Taurus cloud complexes, are placed somewhat farther away in our map than they are generally taken to be. \citet{Zucker2019MolecularCloudsDR2}, which uses a method which is more carefully tailored towards estimating the distance to individual clouds, obtains closer distances to these nearby clouds than does the map presented here, due to the greater number of stars per sightline and different priors used by \citet{Zucker2019MolecularCloudsDR2}.

In Fig. \ref{fig:birdseye-vs-leikeensslin2019}, we compare orthographic projections of our map and LE2019. In Fig. \ref{fig:anticenter-vs-leikeensslin2019}, we compare our map to that of LE2019 in a Sun-centered stereographic projection of the Galactic anticentral region. These comparisons highlight the differences between our methods. As is visible in the orthographic projections, LE2019 produces more spherical dust structures, due partially to the fact that it uses Cartesian voxels, while we use voxels that are elongated in the radial direction. As is visible in the Sun-centered projection, however, our map has much finer angular resolution, due to the much larger number of stars that enter into our analysis. Because we do not require well constrained (or indeed any) Gaia parallaxes in our method, we are able to take into account a far larger number of stars, particularly at larger distances, where Gaia parallaxes become unavailable. The voxels in LE2019 are approximately 2.3~pc on a side, limiting the angular resolution that is achieved by their method, at present. Our map tends to capture more low-reddening dust structure than LE2019, as can be seen, for example, in the bottom right of Fig. \ref{fig:anticenter-vs-leikeensslin2019}. This may be due to the greater number of stars that go into our map, to the quality of our input stellar reddening estimates, or to differences in the choice of priors.

\subsection{Spiral structure} \label{sec:spirals}

\begin{figure*}
    \centering
    \includegraphics[width=\textwidth]{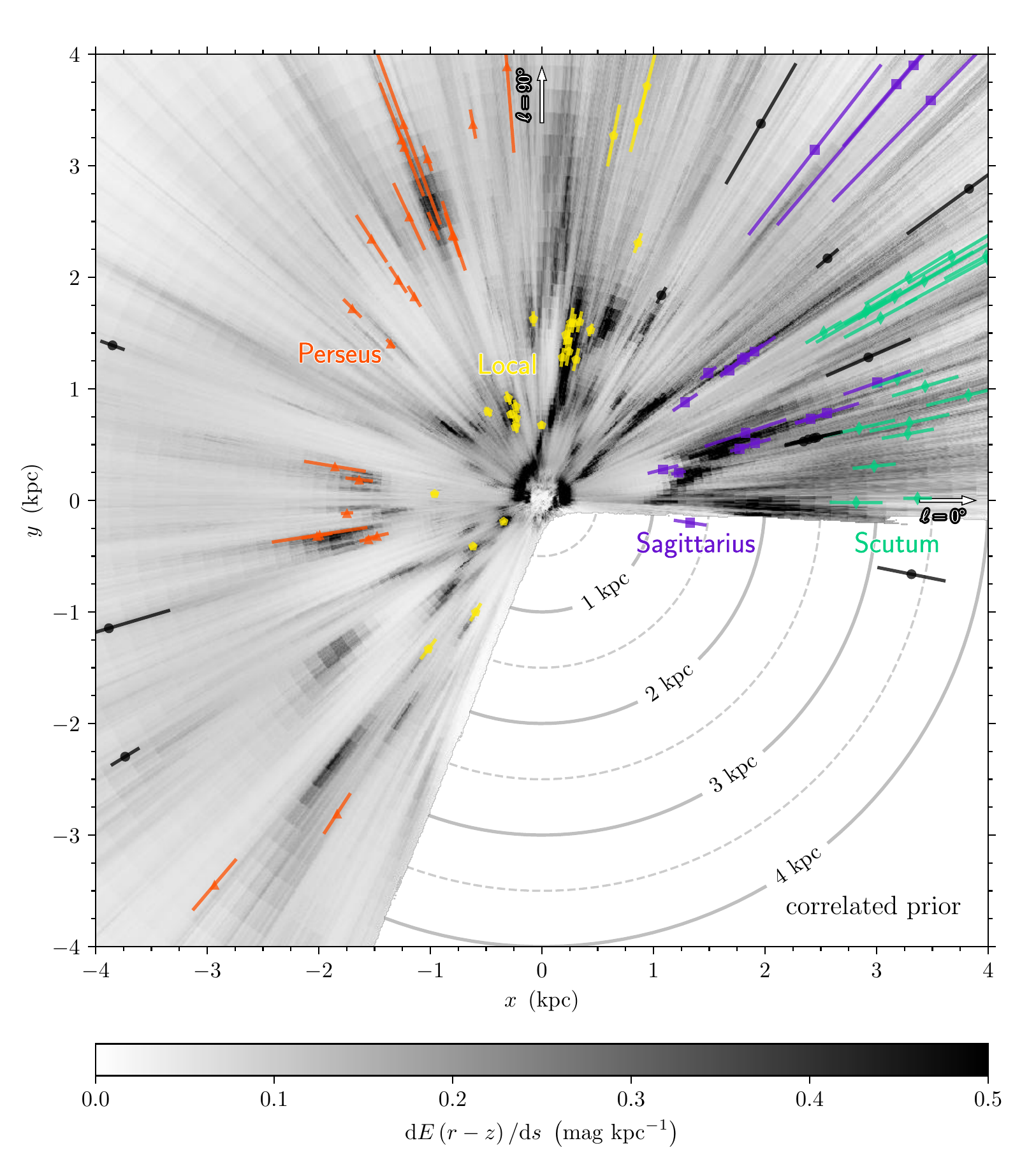}
    \caption{Masers in high-mass star-forming regions \citep{XuReid2016} plotted on top of our inferred dust reddening density. The masers are colored according to the spiral arm that \citet{XuReid2016} assigns them to (the Perseus Arm in orange, the Local Arm in yellow, the Sagittarius Arm in purple and the Scutum Arm in turquoise, all others in black). The distances are calculated naively as the inverse of parallax.}
    \label{fig:masers-on-dust}
\end{figure*}

Widely differing models of the spiral structure of the Milky Way have been proposed, with between two and four arms \citep{GeorgelinGeorgelin1976,Drimmel2000,Vallee2008,ReidMentenZheng2009,HouHan2014}, and with or without a ``molecular ring'' at 4~kpc from the Galactic center \citep{Stecker1975,CohenThaddeus1977,DobbsBurkert2012}. Spiral structure should be apparent not only in 3D maps of the Galaxy, but also in the distribution of interstellar gas and star-forming regions in position-velocity space (see, e.g., Fig. 3 of \citealp{Dame2001}). Though spiral structure is not readily visible in bird's-eye views of our map (see Figs. \ref{fig:birdseye-iterations} and \ref{fig:birdseye-vs-chen2018}), we can test whether the dust in our map preferentially falls along models of spiral arms developed using both position and velocity information.

\citet{XuReid2016} compiles kinematic information for masers in high-mass star-forming regions (HMSFRs), based on very-long-baseline interferometry in the radio, and develops a model of the Milky Way's spiral arms. In Fig. \ref{fig:masers-on-dust}, we compare parallax distances to these HMSFRs with our dust map. Visually, the HMSFRs preferentially lie in regions of higher dust density, as expected. In order to verify this numerically, we compare two Poisson point process models for the distribution of HMSFRs from \citet{XuReid2016}:
\begin{enumerate}
    \item The masers are independently and uniformly distributed in space.
    \item The masers are independently distributed proportionally to the dust reddening density.
\end{enumerate}
We carry out this analysis in a fixed box, defined by $\left| x \right| < 3 \, \mathrm{kpc}$, $\left| y \right| < 3 \, \mathrm{kpc}$, $\left| z \right| < 150 \, \mathrm{pc}$ in a Cartesian representation of Solar-centered Galactic coordinates, excluding regions South of a declination of $-30^{\circ}$, which are not covered by our map. The ratio of the maximum likelihoods of the two models is given by
\begin{align}
    \frac{
        \mathcal{L} \arg{\mathrm{model\ 2}}
    }{
        \mathcal{L} \arg{\mathrm{model\ 1}}
    }
    &=
    \prod_{i=1}^{N_{\mathrm{masers}}}
    \frac{
        \rho \arg{ \vec{r}_i }
    }{
        \left< \rho \right>
    } \, ,
    \label{eqn:maser-likelihood-ratio}
\end{align}
where $\rho \arg{ \vec{r}_i }$ is the dust reddening density (in $\mathrm{mag} \, \mathrm{kpc}^{-1}$) at the location of maser $i$, and $\left< \rho \right>$ is the average reddening density over the entire volume. We find a likelihood ratio of $\sim \exp \arg{27}$, favoring the model in which the masers are distributed according to the reddening density.

We test that this correlation between dust density and maser position is not due to the large-scale distribution of dust and masers (e.g., dust density generally being larger towards the inner Galaxy and HMSFRs also lying preferentially in this direction), but rather correlations between smaller-scale features and maser positions. We do so by adding a small random displacement to the position of each maser ($\pm 300$~pc in $x$ and $y$, and $\pm 10$~pc in $z$), which should reduce any correlations between small-scale features, but keep correlations between large-scale features intact. Comparing the likelihood ratio in Eq. \eqref{eqn:maser-likelihood-ratio} for many realizations of scattered maser positions to the likelihood ratio with the measured maser positions, we find that the measured maser positions lead to a higher ratio in greater than 99\% of cases, indicating that the maser positions are correlated with small-scale features in our map.

Finally, we fit a Poisson point process model to the maser locations in which the masers are distributed according to the density ${\rho \arg{\vec{r}} + \rho_0}$, with $\rho \arg{\vec{r}}$ again being the inferred dust reddening density in our ``median'' 3D dust map, and $\rho_0$ being a free parameter that sets a floor on the maser abundance. The likelihood of this model is maximized when $\rho_0$ satisfies
\begin{align}
    \frac{1}{N} \sum_{i=1}^N \left[ \rho \arg{\vec{r}_i} + \rho_0 \right]^{-1}
    &= \left( \left< \rho \right> + \rho_0 \right)^{-1} \, .
\end{align}
With the \citet{XuReid2016} maser positions in the same volume as used above, we find that ${\rho_0 \approx 0.145 \left< \rho \right>}$. The uniform floor is thus low compared to the average dust density, again indicating that the HMSFRs largely follow the dust density inferred in our 3D dust map.

Despite the correlation between our inferred dust density and the locations of HMSFRs, visual inspection of our dust map does not, in itself, reveal obvious evidence of spiral structure. This could be due to a number of factors, among them:
\begin{enumerate}
    \item Our dust map does not extend beyond a few kiloparsecs. The overall spiral structure of our Galaxy might not be apparent on this scale.
    \item The spiral structure may be less apparent in dust than in the distribution of star-forming regions, particularly on small scales. This is the case in many external spiral galaxies, where the dust density can show complicated structure, with spurs and filaments connecting spiral arms.
\end{enumerate}
Linear features in our map -- for example, the feature running largely along the Local Arm in Fig. \ref{fig:masers-on-dust} -- may correspond to spiral arms and spurs, but a dust map that extends to larger distances will be necessary to see any possible curvature in these features, which one would expect if they correspond to spiral arms. In particular, pushing deeper into the inner Galaxy may allow us to gain a better view of the overall structure of the Milky Way. Doing so will require deeper photometry in the near-infrared than is provided by 2MASS.

\subsection{Stellar Parameter Catalog}
\label{sec:stellar-reddenings}

\begin{figure}
    \centering
    \plottwo{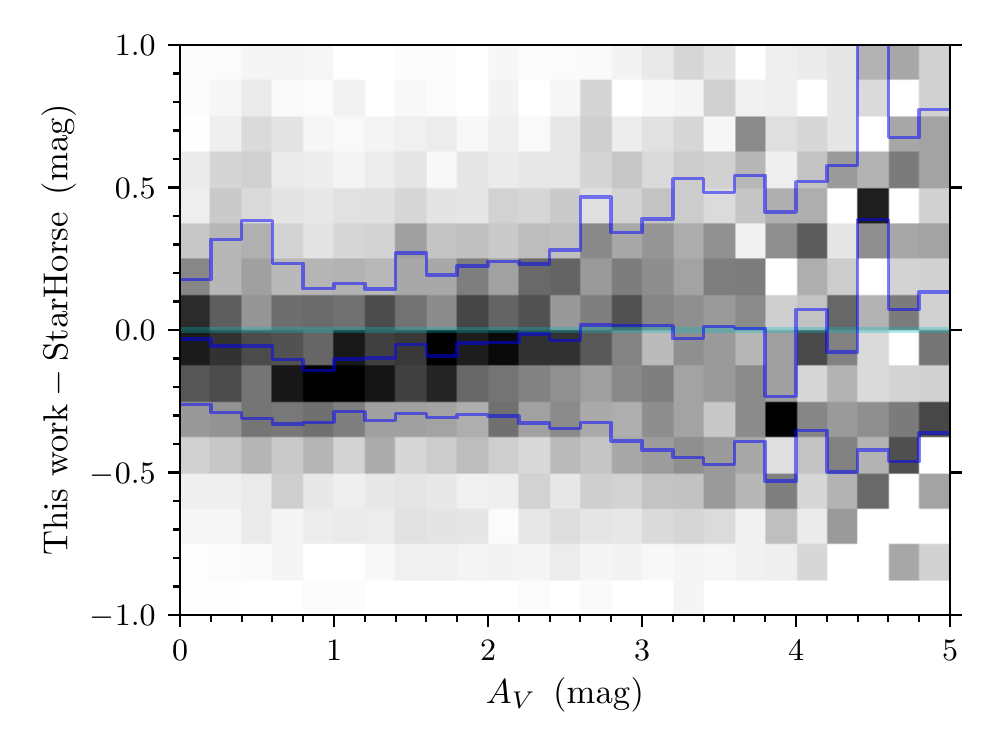}{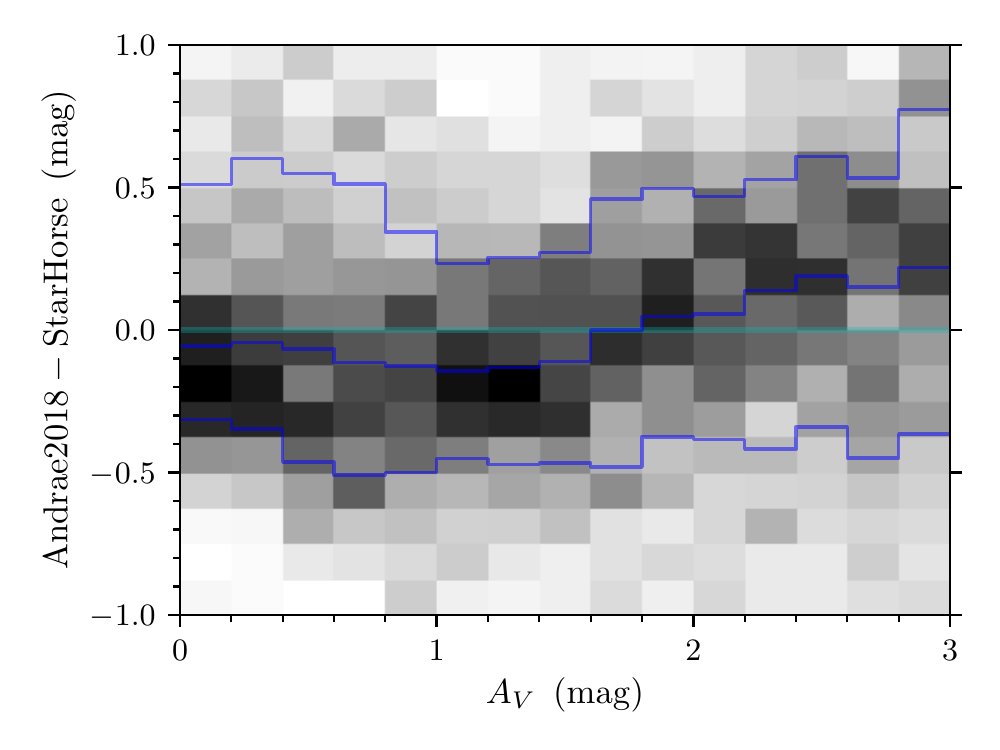}
    \caption{Left panel: Residuals of individual stellar extinctions, in units of $A_V$, obtained by our method and those obtained by StarHorse. The $x$-axis uses the inverse-variance-weighted mean of the two reddening estimates. The blue envelopes show the 16th, 50th and 84th percentiles of the residuals, as a function of $A_V$. Right panel: Comparison between stellar extinctions from Andrae2018 and StarHorse.}
    \label{fig:terra-apsis-vs-starhorse}
\end{figure}

\begin{figure}
    \centering
    \includegraphics[width=0.65\textwidth]{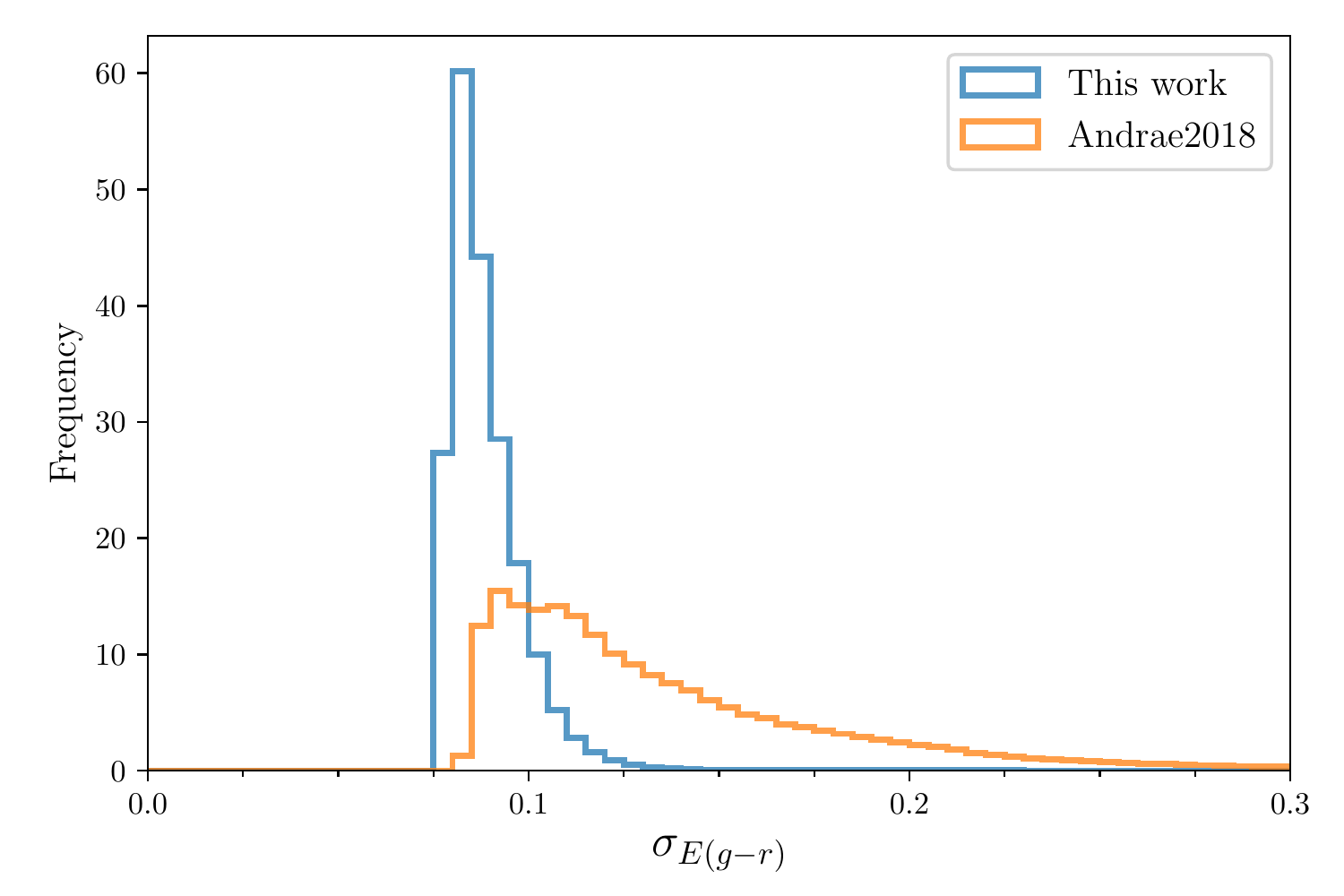}
    \caption{Histogram of uncertainties in the stellar reddenings inferred by our method and by Andrae2018. These uncertainties include the systematic error floors determined by comparison with extinctions inferred by StarHorse.}
    \label{fig:hist-terra-andrae2018-errors}
\end{figure}

\begin{figure}
    \centering
    \includegraphics[width=\textwidth]{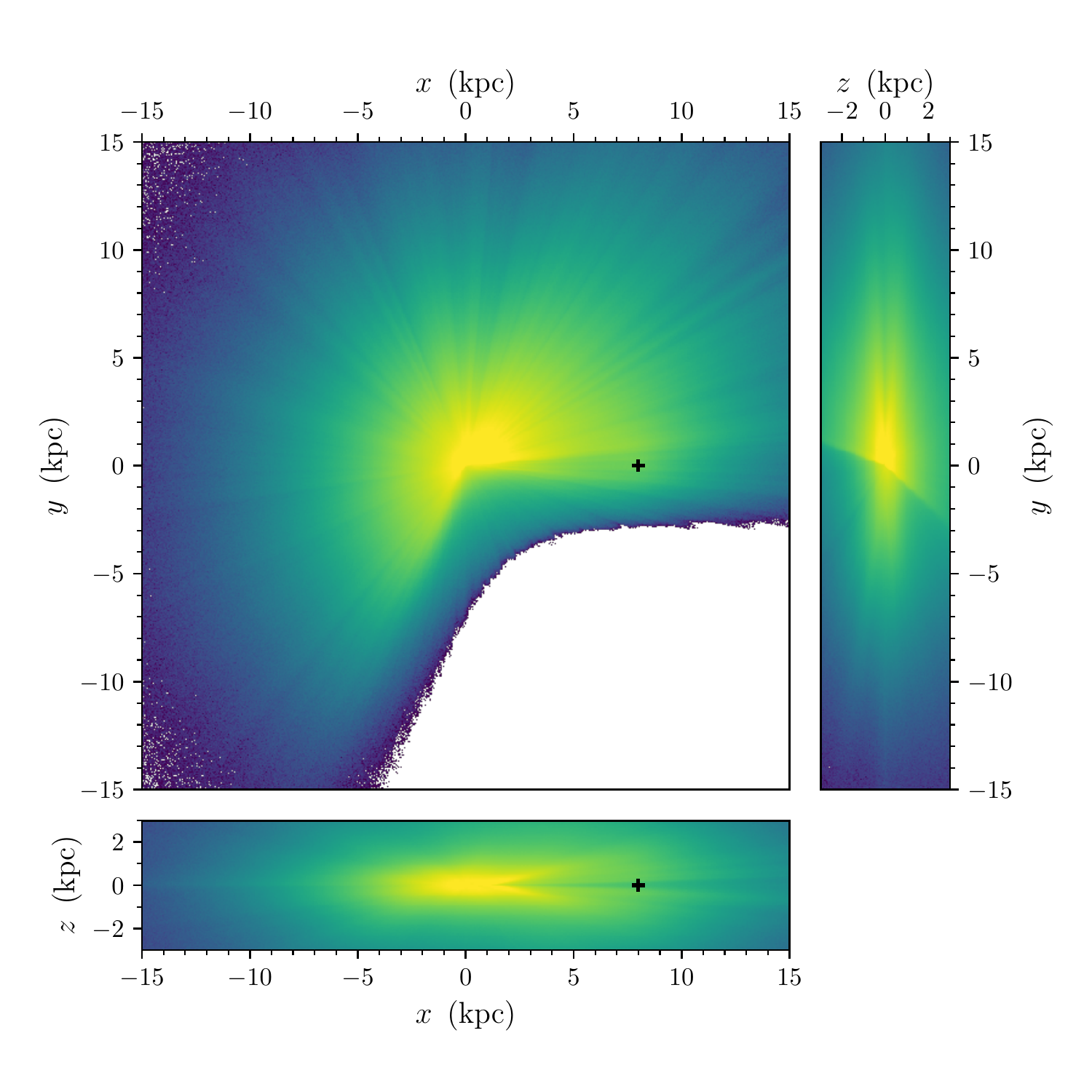}
    \caption{Distribution of stars in our catalog, in Sun-centered Cartesian coordinates. Each panel shows a projection of the number density of stars in our catalog, on an arbitrary logarithmic scale. The Galactic center is marked by a cross in the $\left( x,\,y \right)-$ and $\left( x,\,z \right)$-projections.}
    \label{fig:stellar-number-density}
\end{figure}

\begin{figure}
    \centering
    \includegraphics[width=\textwidth]{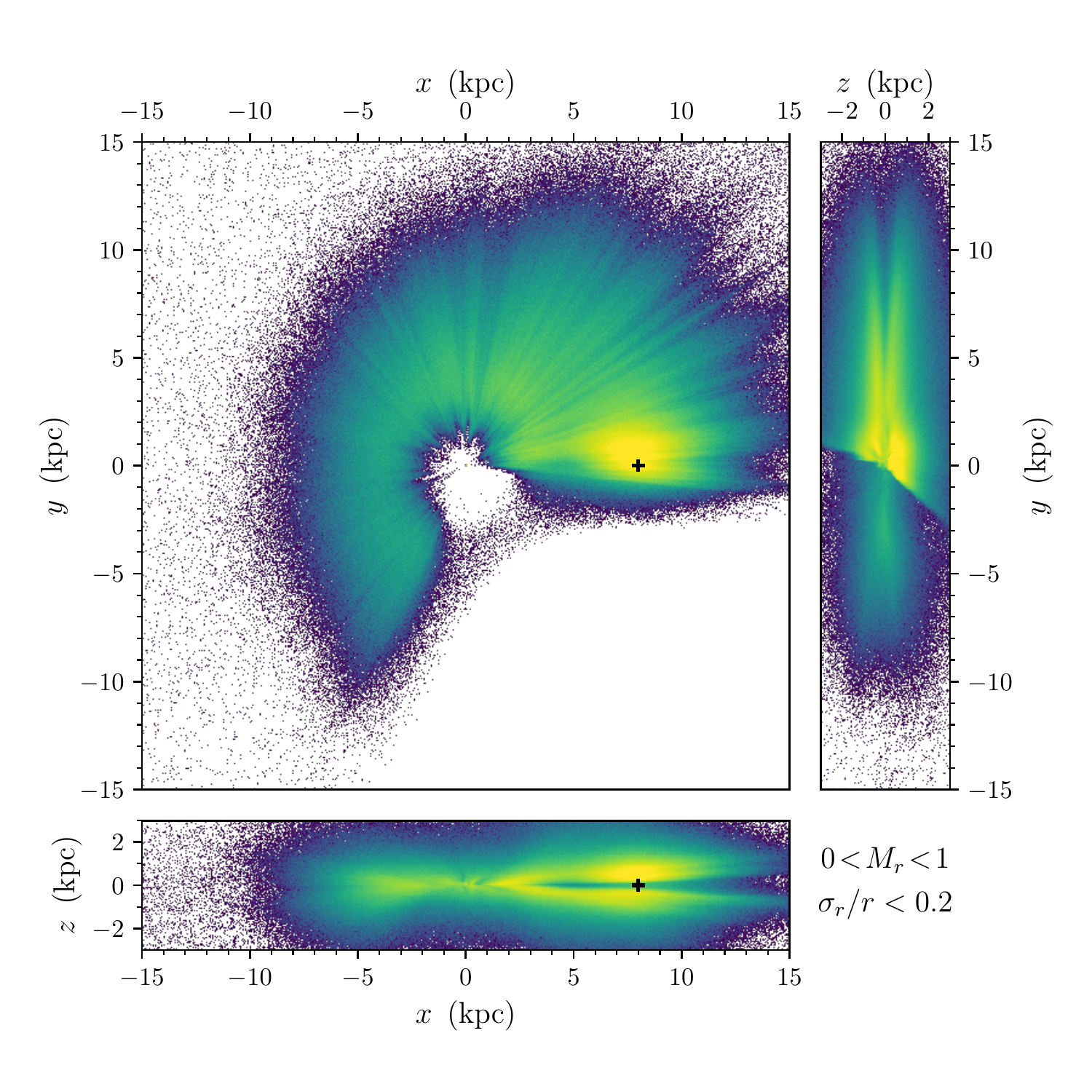}
    \caption{Distribution of stars of absolute magnitude $0 < M_r < 1$ with well-determined distances ($\nicefrac{\sigma_r}{r} < 0.2$) in our catalog, in Sun-centered Cartesian coordinates. Each panel shows a projection of the stellar number density, on an arbitrary logarithmic scale. For these relatively bright stars, the Galactic bulge is apparent as a region of increased number density near the Galactic center (which is marked by a cross).}
    \label{fig:stellar-number-density-selection}
\end{figure}

Inferring stellar distances, reddenings and types (absolute $r_{\mathrm{P1}}$-band magnitude and $\left[ \mathrm{Fe} / \mathrm{H} \right]$) is a necessary first step towards generating a 3D map of reddening, as described in Section \ref{sec:single-star}. We save samples of these stellar parameters, and make them available along with our 3D reddening map. In order to validate the accuracy of our stellar reddening estimates, we compare them to those estimated by \citet{Queiroz2018StarHorse} (hereafter, ``StarHorse''), which are based on APOGEE DR14 spectroscopy \citep{Abolfathi2018APOGEEDR14}, 2MASS $JHK_s$ photometry, APASS DR9 $BVg^{\prime}r^{\prime}i^{\prime}$ photometry \citep{Henden2014APASS} and Gaia DR1 parallaxes and $G$-band photometry. We also compare our stellar reddenings to those of \citet{Andrae2018DR2Apsis} (hereafter, ``Andrae2018''), which are based on Gaia DR2 parallaxes and $G$-band, BP and RP photometry. StarHorse estimates $A_V$, while Andrae2018 estimates both $A_G$ and $\EBPRP$. Here, we compare our reddenings to $\EBPRP$.

We first match our stellar reddening catalog to that of Andrae2018 using Gaia DR2 \texttt{source\_id}, excluding stars for which the best-fit $\chi^2 / \mathrm{passband} > 5$ for our stellar model. This yields a catalog of 41.3 million sources. We convert our reddening estimates, which are in the arbitrary unit $E$, to $\Egr$ using the relation
\begin{align}
    \Egr = \left( 0.901 \, \mathrm{mag} \right) E \, ,
\end{align}
as implied by the extinction coefficients in Table \ref{tab:extinction-vector}. We then match our combined catalog to the StarHorse APOGEE DR14 catalog using a matching radius of $0.2^{\prime \prime}$, obtaining a catalog of 5350 stars.

In order to compare these catalogs on a uniform scale, we have to convert them to the same unit of reddening or extinction. In order to do this, we fit a linear relationship between the StarHorse $A_V$ estimates and our $\Egr$ estimates, and a separate linear relationship between StarHorse $A_V$ and the Andrae2018 estimates of $\EBPRP$. Similarly to the method described in \citet{HoggBovyLang2010}, we also fit nuisance parameters describing whether each star is an outlier and the overall distribution of outliers. We additionally fit a systematic error floor for our $\Egr$ estimates and for Andrae2018's $\EBPRP$ estimates, assuming that the uncertainties reported by StarHorse are accurately estimated. The details of this model are described in Appendix \ref{app:linear-model}.

We obtain the relations
\begin{align}
    \Egr &= 0.269 \, A_{V, \, \mathrm{StarHorse}} + 0.03 \, \mathrm{mag} \, , \\
    \EBPRP &= 0.294 \, A_{V, \, \mathrm{StarHorse}} + 0.11 \, \mathrm{mag} \, ,
\end{align}
with a systematic error floor of 0.08~mag on our $\Egr$ estimates, and a systematic error floor of 0.09~mag on the Andrae2018 $\EBPRP$ estimates. With these relations, we are able to transform between the three reddening measurements, and compare them on a common scale. The left panel of Fig. \ref{fig:terra-apsis-vs-starhorse} shows the residuals stellar $A_V$ recovered by our method and by StarHorse, as a function of $A_V$, while the right panel shows the comparison between Andrae2018 and StarHorse (likewise in $A_V$). As StarHorse makes use of spectroscopically determined stellar atmospheric parameters, its extinction estimates are generally much more precise than those produced by our method or by Andrae2018. The residuals in Fig. \ref{fig:terra-apsis-vs-starhorse} are thus dominated by errors in our measurements (in the left panel) or in Andrae2018's measurements (in the right panel). As can be seen from these panels, the errors in our stellar extinction estimates are generally smaller than those of Andrae2018, due to the greater number of photometric passbands used in our study (up to eight, versus two). In Fig. \ref{fig:hist-terra-andrae2018-errors}, we directly compare the uncertainties in stellar reddening inferred by our method and Andrae2018, for the 41.3 sources in our matched catalog, with the recovered systematic error floors added in quadrature. In $\Egr$, the $95^{\mathrm{th}}$ percentile of the reddening uncertainties is 0.11~mag for our estimates, and 0.27~mag for the Andrae2018 estimates. In this matched catalog, our median uncertainty is 0.086~mag, 30\% less than the median uncertainty of 0.124~mag in Andrae2018. These represent the typical uncertainties that can be expected for relatively bright stars with well-measured Gaia parallaxes. However, we obtain individual stellar reddening estimates for all 799~million stars in our input catalog, including those without Gaia parallax measurements. We make this entire catalog available to the community, as described in Section \ref{sec:access}.

Figs. \ref{fig:stellar-number-density} shows the spatial distribution of stars in our catalog, excluding those with poor goodness-of-fit (maximum-likelihood ${\chi^2 / \mathrm{passband} > 5}$). This distribution is determined by the true number density of stars throughout the Galaxy, the sensitivities of the PS1, 2MASS and Gaia surveys, our selection function, and the distribution of interstellar dust. With a good understanding of the latter confounding effects, this catalog can be used as a basis for determining the true distribution of stars throughout the Galaxy. Fig. \ref{fig:stellar-number-density-selection} shows the distribution of stars with absolute $\rps$ magnitudes ${0 < M_r < 1}$ and distances that are determined to better than 20\%. Although a detailed treatment of selection functions is necessary to determine the structure of the Galaxy, the Galactic bulge is visible as in increase in stellar density near the Galactic center in the projection to the $\left( x, \, y \right)$ plane. Because the boundary of the PS1 survey passes essentially through the Galactic center, the bulge appears off-center from the Galactic center. The relative lack of stars directly in the Galactic midplane (${z = 0}$) is due to dust obscuration, while the lack of bright stars close to the Sun is due to saturation in photometric surveys. This catalog will serve as a basis for future work on the structure of our Galaxy.

\section{Discussion} \label{sec:discussion}

The work presented here can be extended in a number of ways, both through improvements of the model and incorporation of more data. One methodological improvement would be to optimize the kernel parameters in the spatial prior. In this work, we have demonstrated that the incorporation of a Gaussian process prior on the logarithm of the dust reddening density yields significant improvements in the effective distance resolution of the map, with cloud distances becoming much more precise. However, the kernel that we impose is determined beforehand, rather than inferred from the data. An optimized kernel would allow us to extract more information about the dust reddening density from our data. As \citet{LeikeEnsslin2019} shows, with an optimized kernel, it is possible to extract a significant amount of information about the spatial distribution of dust from even relatively small stellar catalogs (3.7 million stars, albeit with well determined parallaxes, versus $\sim$800 million in this work). An additional improvement to the method here would be to take into account larger patches of sky surrounding each sightline, in order to better approximate the target Gaussian process prior. This is particularly important in the near regime, in which our angular patches correspond to smaller physical extent. Finally, although we increase the number of distance bins in this work by a factor of four over \citet{Green2018}, our voxels are still elongated in the radial direction. This corresponds to the fundamental fact that we have much better knowledge of the angular distribution of dust than of its distribution with distance, but it makes implementation of spatially correlated priors -- particularly those that obey the Copernican principle -- more difficult.

A separate area in which significant improvement is achievable is in the dust reddening curve that we assume. At present, our work (like all other 3D dust maps that we are aware of) assumes a single, universal dust extinction curve for the entire Galaxy. However, the shape of the dust extinction spectrum varies significantly throughout the Galaxy \citep{FitzpatrickMassa1986ExtinctionI,FitzpatrickMassa1988ExtinctionII,FitzpatrickMassa1990ExtinctionIII,FitzpatrickMassa2005ExtinctionIV,FitzpatrickMassa2007ExtinctionV,FitzpatrickMassa2009ExtinctionVI}. Although there are multiple features in the dust extinction spectrum which can vary independently (such as the width and strength of the 2175\,\r{A} bump, the slope of the wavelength-extinction relation in the optical, and the shape of the rise in extinction in the far ultraviolet), the dust extinction spectrum is often parameterized as a single-parameter family of curves \citep{Cardelli1989RV}, using the ratio of total to selective extinction, $R_V$, defined as:
\begin{align}
    A_V = R_V \, E \arg{B-V} \, .
    \label{eqn:RV-definition}
\end{align}
In the diffuse interstellar medium of the Milky Way, $R_V$ is typically taken to be approximately 3.1 \citep{Cardelli1989RV}. However, the value of $R_V$ can vary significantly throughout the interstellar medium. \citet{Schlafly2017MappingExtinctionCurve} combines $R_V$ determinations for individual stars \citep{Schlafly2016ExtinctionVector} with the Bayestar2015 3D dust map \citep{Green2015} to map the spatial variation of the extinction curve throughout the Milky Way, finding variations on kiloparsec scales. As most dust maps that use stellar photometry rely primarily on changes in stellar colors, and are less sensitive to changes in absolute magnitude, they are essentially maps of dust reddening, rather than extinction. Extinction maps derived from these reddening measurements (e.g., using Eq. \ref{eqn:RV-definition}) without knowledge of the spatial dependence of $R_V$ will have large-scale, spatially dependent systematic errors. Tackling this source of systematics is of particular importance to precision cosmological measurements \citep{Nataf2016RVCosmology,Huterer2013}.

One way to address variation in $R_V$, without unduly complicating our model, is to treat the dust reddening density as a pair of numbers at each point in space. \citet{Schlafly2016ExtinctionVector} decomposes the reddening vector into a mean component, and an orthogonal vector along which most of the variation in the reddening vector occurs. The virtue of this formulation, from the perspective of 3D dust mapping, is that the two components of reddening add linearly, simplifying their inclusion in our model. \citet{Zucker2019MolecularCloudsDR2}, which measures distances to a catalog of molecular clouds using stellar photometry and Gaia parallaxes, takes advantage of this linearity to determine the distance and $R_V$ of the dust.

More generally, any systematics in reddening maps at high Galactic latitude are of importance to cosmology. Both this work and \citet{Green2018} have identified systematic trends in reddening when comparing with the Planck14 and SFD dust maps, particularly for reddenings of ${E \arg{g-r} \lesssim 0.1 \, \mathrm{mag}}$. These systematics could be due to the uncertain zero-point of the dust reddening (e.g., what is the absolute reddening of some reference point on the sky?) and variation in $R_V$ across the high-Galactic-latitude sky. An additional worry for cosmology is contamination from large-scale extra-Galactic structure. For far-infrared emission-based maps, such as Planck14 and SFD, this contamination comes from thermal dust emission in external galaxies residing at a wide range in redshift \citep{ChiangMenard2019}. For maps based on stellar colors, such as ours, this contamination comes from quasars and unresolved galaxies misconstrued as stars \citep{ChiangMenard2019}. More effective rejection of these classes of objects (for example, by treating them explicitly in our model) will be important to providing reddening maps for use in precision cosmological measurements.

There are also additional datasets which can be applied to the dust-mapping problem. Deeper near-infrared photometry, in particular, will allow us to trace dust to greater extinctions. Dust maps which extend to greater distances, particularly in the Galactic midplane, will allow us to better uncover the overall structure of the Milky Way. As discussed in Section \ref{sec:spirals}, with our present map, extending only a few kiloparsecs, the spiral structure of the Galaxy is not readily apparent. Incorporation of deep infrared photometry -- e.g., from the Spitzer GLIMPSE surveys \citep{Churchwell2009GLIMPSE}, the ongoing UKIDSS Galactic Plane Survey \citep{Lucas2008UKIDSSGPS} in the North and the VISTA Variables in the Via Lactea survey in the South \citep{Minniti2010VVV}, and from the newly published unWISE catalog \citep{Schlafly2019unWISE} -- would extend the range of our dust map, particularly in the inner Galaxy. This would allow us to directly see the overall structure of the spiral arms, rather than drawing curves through short segments of them, as we are currently limited to doing with available 3D dust maps.

Finally, the work presented in this paper is limited to the Galaxy north of a declination of $-30^{\circ}$. The ongoing Dark Energy Camera Plane Survey \citep[][hereafter ``DECaPS'']{Schlafly2018DECaPS} has already filled in the Southern Galactic plane in the region $\left| b \right| \lesssim 5^{\circ}$, and is in the process of being extended to $\left| b \right| \lesssim 10^{\circ}$. DECaPS observes in five PS1-like passbands on the Dark Energy Camera, mounted on the Blanco 4-m telescope in Cerro Tololo, Chile. Incorporation of the resulting photometry will allow us to map all $360^{\circ}$ of the Galactic midplane.

\section{Accessing the map}
\label{sec:access}

The dust map presented here can be downloaded at \href{https://doi.org/10.7910/DVN/2EJ9TX}{doi:10.7910/DVN/2EJ9TX}. The easiest way to interact with the map, however, is through the Python package \texttt{dustmaps} \citep{Green2018dustmaps}, which provides a uniform interface to a range of maps of dust reddening and extinction, including all of the dust maps discussed in this paper. The map can additionally be queried interactively at \href{http://argonaut.skymaps.info}{argonaut.skymaps.info}.

In addition to the dust map, we provide our individual stellar inferences (described in Section \ref{sec:single-star}). Samples of reddening, distance modulus, absolute $\rps$ magnitude and metallicity for 799 million stars can be downloaded at \href{https://doi.org/10.7910/DVN/AV9GXO}{doi:10.7910/DVN/AV9GXO}. This is an order of magnitude more stellar reddenings than are provided by the Gaia DR2 catalog \citep{Andrae2018DR2Apsis}. Our technique leverages a greater number of photometric passbands (between four and eight, compared to two independent passbands for Gaia DR2), and delivers typical reddening uncertainties that are 30\% lower. For stars which are present in Gaia DR2, we provide a cross-match to the corresponding Gaia DR2 \texttt{source\_id}.

\section{Conclusion}
\label{sec:conclusion}

We have presented a new 3D map of dust reddening, covering three quarters of the sky out to a distance of several kiloparsecs. This map is based on stellar distances and reddenings, inferred from Gaia DR2 parallaxes and optical and near-infrared photometry from PS1 and 2MASS. This dust map has four times the distance resolution as \citet{Green2018}. Another improvement over our previous method is that we impose a Gaussian-process prior on the logarithm of the dust reddening density, using an approximation that requires far less computational time than a naive approach would require. This prior encodes our knowledge that the spatial distribution of dust should be smooth on some scale, and has the effect of sharpening our distance determinations to dust clouds, particularly in the nearest kiloparsec, and reducing noise in our reddening estimates.

We have made both the 3D dust map, and the parameter inferences (distances, reddenings and types) for 799 million stars that underlie the map, available at \href{https://doi.org/10.7910/DVN/2EJ9TX}{doi:10.7910/DVN/2EJ9TX} and \href{https://doi.org/10.7910/DVN/AV9GXO}{doi:10.7910/DVN/AV9GXO}, respectively. The dust map is also available through the Python package \texttt{dustmaps} \citep{Green2018dustmaps}. For the stellar parameter inferences, we provide matches to the Gaia DR2 catalog.

At high Galactic latitude, our integrated reddening estimates agree well with those of Planck14 and Bayestar17. We find median residuals between our map and those of the far-infrared radiance-based Planck14 reddening map of ${\Delta E \arg{g-r} \lesssim 0.02 \, \mathrm{mag}}$, out to a reddening of ${E \arg{g-r} \approx 1 \, \mathrm{mag}}$. Our new map matches Planck14 better at low column densities, due to the superior signal-to-noise ratio of its reddening determinations.

Comparisons with other 3D dust maps show similar overall structures, though the maps have different resolutions and ranges of validity. We see the same basic features as Chen19 within a distance of $\lesssim 2 \, \mathrm{kpc}$, although we trace the dust distribution with finer distance resolution and to greater distances. Much of this difference is likely attributable to our use of deeper optical photometry from PS1, as well as the use of a Gaussian-process prior. In the nearby regime, in a $\left( 600\,\mathrm{pc} \right)^3$ box centered on the Sun, we compare our map with that of LE2019. There are many differences between our methods, with LE2019 employing a Cartesian voxelization of the dust, in contrast to the spherical voxelization that we choose. LE2019 likewise imposes a Gaussian-process prior on the logarithm of the dust reddening density, but infers the parameters of the dust reddening correlation function as part of a hierarchical model. The spatial resolution of the resulting LE2019 map is greater than ours, while the angular resolution of our map is greater.

The spiral structure of the Milky Way galaxy is uncertain, with a wide range of models having been developed. \citet{XuReid2016} uses HMSFRs with kinematic and distance measurements from embedded masers to trace spiral arms. We have shown that the masers compiled by \citet{XuReid2016} are preferentially distributed in regions of high reddening density in our map. However, in order to use the 3D distribution of dust reddening to assess models of spiral arms, it is necessary to extend our map beyond the current limits of a few kiloparsecs, particularly in the inner Galaxy. One way of achieving this will be the incorporation of deeper near-infrared photometry, allowing us to see stars through far greater dust column densities.

\section{Acknowledgements}
\label{sec:acknowledgements}

The computations in this paper were run on the Odyssey cluster supported by the FAS Division of Science, Research Computing Group at Harvard University.

This work took part under the program Milky-Way-Gaia of the PSI2 project funded by the IDEX Paris-Saclay, ANR-11-IDEX-0003-02.

E.S. acknowledges support for this work by NASA through ADAP grant NNH17AE75I and Hubble Fellowship grant HST-HF2-51367.001-A awarded by the Space Telescope Science Institute, which is operated by the Association of Universities for Research in Astronomy, Inc., for NASA, under contract NAS 5-26555. D.F. and C.Z. acknowledge support by NSF grant AST-1614941, ``Exploring the Galaxy: 3-Dimensional Structure and Stellar Streams.''

This work has made use of data from the European Space Agency (ESA) mission
{\it Gaia} (\url{https://www.cosmos.esa.int/gaia}), processed by the {\it Gaia}
Data Processing and Analysis Consortium (DPAC,
\url{https://www.cosmos.esa.int/web/gaia/dpac/consortium}). Funding for the DPAC
has been provided by national institutions, in particular the institutions
participating in the {\it Gaia} Multilateral Agreement.

The Pan-STARRS1 Surveys (PS1) have been made possible through contributions of the Institute for Astronomy, the University of Hawaii, the Pan-STARRS Project Office, the Max-Planck Society and its participating institutes, the Max Planck Institute for Astronomy, Heidelberg, and the Max Planck Institute for Extraterrestrial Physics, Garching, The Johns Hopkins University, Durham University, the University of Edinburgh, Queen's University Belfast, the Harvard-Smithsonian Center for Astrophysics, the Las Cumbres Observatory Global Telescope Network Incorporated, the National Central University of Taiwan, the Space Telescope Science Institute, the National Aeronautics and Space Administration under Grant No. NNX08AR22G issued through the Planetary Science Division of the NASA Science Mission Directorate, the National Science Foundation under grant No. AST-1238877, the University of Maryland, and E\"{o}tv\"{o}s Lor\'{a}nd University (ELTE).

This publication makes use of data products from the 2MASS, which is a joint project of the University of Massachusetts and the Infrared Processing and Analysis Center/California Institute of Technology, funded by the National Aeronautics and Space Administration and the National Science Foundation.

\software{
    Astropy \citep{astropy2013,astropy2018},
    dustmaps \citep{Green2018dustmaps},
    emcee \citep{ForemanMackey2013emcee},
    Matplotlib \citep{Hunter2007Matplotlib},
    NumPy \citep{NumPy},
    SciPy \citep{SciPy}
}

\appendix

\section{Grid-evaluation of stellar posterior densities} \label{app:grid-stars}

In contrast to our approach in \citet{Green2014,Green2015} and \citet{Green2018}, here we brute-force evaluate the single-star posterior probability densities on a regular grid, rather than approximating them using a kernel density estimate on Markov Chain Monte Carlo samples. This brute-force approach is feasible -- and indeed extremely fast -- with a few approximations. For any given stellar type, the observed magnitudes depend only linearly on the reddening and distance modulus. As the uncertainties on our observed magnitudes are Gaussian, the likelihood of any parameters that the magnitudes depend on linearly is also Gaussian. In other words,
\begin{align}
    \pcond{\hat{m}}{\Theta, \, \mu, \, E}
\end{align}
is Gaussian in $\mu$ and $E$ for fixed $\Theta$. Moreover, the covariance of this Gaussian depends only on the observational uncertainties and the reddening curve.

Our scheme for calculating the stellar posteriors is thus as follows:
\begin{enumerate}
    \item For each point in a uniform grid in $\Theta$, calculate the $\mu$ and $E$ that maximize the likelihood $\pcond{\hat{m}}{\Theta, \, \mu, \, E}$. Call these values $\mu^{\ast}$ and $E^{\ast}$.
    \item Add a Gaussian centered on each $\mu^{\ast}$ and $E^{\ast}$, weighted by $\pcond{\hat{m} , \, \hat{\varpi}}{\Theta, \, \mu^{\ast}, \, E^{\ast}} p \arg{\Theta, \, \mu^{\ast}, \, E^{\ast}}$.
\end{enumerate}
The second step makes the approximation that over the Gaussian centered on a given $\left( \mu^{\ast}, \, E^{\ast} \right)$, the likelihood in parallax and the prior on $\mu$ are constant. In our testing, this approximation typically only had a small effect.

Using this method, we evaluate $\pcond{\mu, \, E}{\hat{m}, \, \hat{\varpi}}$ over a grid in distance modulus and reddening, with a bin scale of 0.125~mag in distance modulus and 0.01~mag in reddening. This information is carried over into the line-of-sight inference. However, in order to save disk space, we ultimately store samples from the posterior, discarding the grid-evaluated posterior density after we have completed the line-of-sight inference.

This scheme is fast. On a single core, with a library of $\sim$41000 stellar templates, we are typically able to evaluate the gridded posterior densities of $\sim$50 stars per second.

\section{Round-robin sampling of neighboring pixels}
\label{app:neighbor-round-robin}

To update each neighboring sightline, we choose a new sample from the previous iteration, conditioned on the reddening densities in the other pixels in the patch. This is essentially an importance-sampling step, and we must choose the new sample according to the ratio of the new posterior density to the posterior density of the sample in the previous iteration. During each iteration, we must therefore store the posterior density for each sample of the central pixel, to be used when resampling pixels in the following iterations.

How does one calculate this posterior density? For a given sightline $i$,
\begin{align}
    \pcond{\alpha_i}{\hat{m}_i, \, \hat{m}_{\backslash i}}
    &= \int \pcond{\alpha_i, \, \alpha_{\backslash i}}{\hat{m}_i, \, \hat{m}_{\backslash i}} \mathrm{d} \alpha_{\backslash i} \\
    &= \int \pcond{\alpha_i}{\hat{m}_i, \, \hat{m}_{\backslash i}, \, \alpha_{\backslash i}} \pcond{\alpha_{\backslash i}}{\hat{m}_i, \, \hat{m}_{\backslash i}} \mathrm{d} \alpha_{\backslash i} \\
    &=
    \frac{1}{\mathcal{Z}} \,
    \pcond{\hat{m}_i}{\alpha_i}
    \int
    \pcond{\alpha_i}{\alpha_{\backslash i}}
    \pcond{\alpha_{\backslash i}}{\hat{m}_i, \, \hat{m}_{\backslash i}}
    \mathrm{d} \alpha_{\backslash i}
    \\
    &=
    \frac{1}{\mathcal{Z}} \,
    \underbrace{
        \pcond{\hat{m}_i}{\alpha_i}
    }_{
        \equiv \mathcal{L}_i
    }
    \underbrace{
        \left<
            \pcond{
                \alpha_i
            }{
                \alpha_{\backslash i}
            }
        \right>_{
            \alpha_{\backslash i}
            |
            \hat{m}_i, \, \hat{m}_{\backslash i}
        }
    }_{
        \equiv \mathcal{P}_i
    }
    \, ,
    \label{eqn:central-posterior-density-integral}
\end{align}
where $\mathcal{Z}$ is a normalizing constant dependent only on the data, and can therefore be ignored. For simplicity, $\hat{m}$ here denotes both photometry and parallaxes. The second term, $\mathcal{L}_i$, is the likelihood of $\alpha_i$, which does not change from iteration to iteration. The third term, $\mathcal{P}_i$, is the ``effective prior'' on $\alpha_i$ generated by the neighboring sightlines, averaged over the range of values that the reddening in those sightlines can take.

In the first iteration, the calculation of this ``effective prior,'' $\mathcal{P}_i$, is trivial. As the pixels are independent, it is simply equal to the prior on $\alpha_i$. However, in subsequent iterations, we must estimate $\mathcal{P}_i$ for each sample we store of $\alpha_i$, by calculating the average of $\pcond{\alpha_i}{\alpha_{\backslash i}}$ over samples of $\alpha_{\backslash i}$ drawn from the chain. This is done as a post-processing step after we are done sampling the patch of sky. The computational complexity of this step is $\mathcal{O} \arg{n^2}$ in the number of samples stored, because for each of the $n$ samples of $\alpha_i$ we store, we have to average over $\pcond{\alpha_i}{\alpha_{\backslash i}}$ for $n$ samples of $\alpha_{\backslash i}$.

In order to select a new sample for one of the neighboring pixels $j$, we therefore select one of the stored samples from the previous iteration, with the probability of selecting sample $k$ being proportional to
\begin{align}
    w_j^k
    &=
    \frac{
        \pcond{\hat{m}_j}{\alpha_j^k} \pcond{\alpha_{j}^k}{\alpha_{\backslash j}}
    }{
        \mathcal{L}_j^k \mathcal{P}_j^k
    }
    =
    \frac{
        \pcond{\alpha_{j}^k}{\alpha_{\backslash j}}
    }{
        \mathcal{P}_j^k
    }
    \, ,
\end{align}
where $\mathcal{L}_j^k$ and $\mathcal{P}_j^k$ are the stored likelihood and effective prior, respectively, for sample $k$ of sightline $j$ from the previous iteration.

If the stored samples for the neighboring pixels are a decent approximation to the new correlated posterior on neighboring sightlines, then this sampling procedure will work well. Over succeeding iterations, we slowly change the covariance kernel, from completely diagonal (i.e., independent sightlines) to the target kernel. If we change the covariance kernel too drastically from one iteration to the next, then the stored samples will be a poor representation of the posterior density, and will not be able to represent the state of the neighboring sightlines in the current iteration. As in importance sampling, this would manifest itself as the entropy of the weight distribution $\left\{ w_j^k \, | \, k = 1, \, 2, \, \ldots, n \right\}$ approaching zero.

\section{Improving Markov Chain mixing with the ``deformation ladder''}
\label{app:ladder}

\begin{figure}
    \centering
    \includegraphics[width=0.45\textwidth]{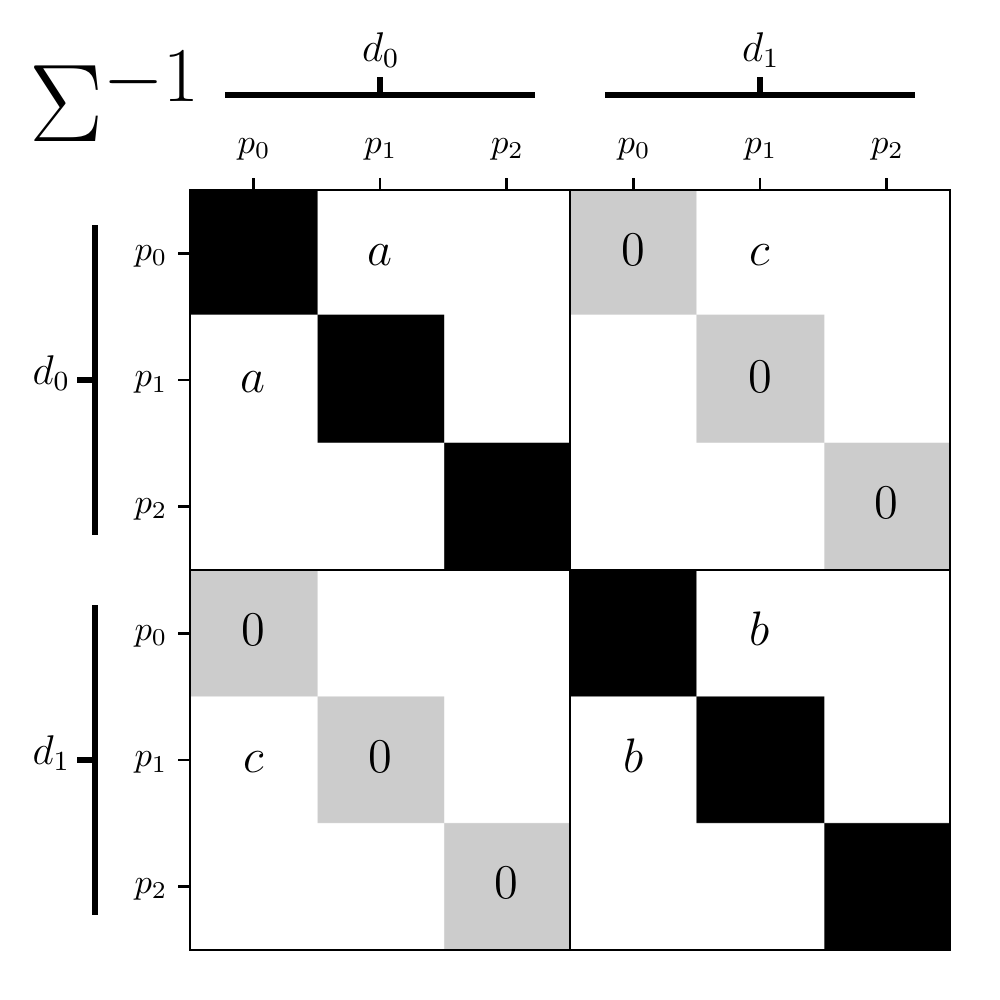}
    \caption{Deformation of the inverse covariance matrix. The entries corresponding to different angular pixels ($p_0$, $p_1$, etc.) and distance bins ($d_0$, $d_1$, etc.) are labeled. In the original inverse covariance matrix, the cross-distance entries are zero. As we deform the inverse covariance matrix, we modify the entries linking \textit{different} angular pixels in neighboring distance bins. For example, the entry $c$ is be set to $\frac{\epsilon}{2} \left( a + b \right)$. We do not alter entries linking the \textit{same} angular pixel in neighboring distance bins. Afterwards, the entire matrix is multiplied by the scalar $\left( 1 + 2 \epsilon \right)^{-1}$. The variable $\epsilon$ controls the amount of deformation, with $\epsilon = 0$ leaving the matrix unaltered.}
    \label{fig:icov-deformation}
\end{figure}

The sampling procedure for a patch of sky described above can suffer from poor Markov Chain mixing. Our distance bins are significantly longer in the radial than in the angular direction (by a typical factor of $\sim$30), leading to the transverse correlations between neighboring voxels at the same distance being much greater than correlations between pixels at different distances. One possible failure mode of our sampling procedure may occur when there is a dust cloud at a particular distance, spanning several neighboring sightlines. The correlated prior will tend to favor solutions in which the dust is placed at the same distance in all the sightlines. The prior imposes a penalty if the dust cloud is moved forward or backward in distance in just one sightline. With the round-robin sampling procedure we employ for each patch of sky, this means that it is difficult for the cloud to transition from one distance to another. The sampler becomes locked in a state with the cloud at one distance, and has to traverse a high potential barrier in order to reach solutions with the cloud at different distances.

In order to deal with this problem, we employ a method which is conceptually similar to parallel tempering. Parallel tempering works by creating a ladder of samplers, with the bottom rung sampling from the target distribution, and successively higher rungs sampling from progressively flattened versions of the distribution (using a sampling temperature parameter to control the flattening of the distribution). The higher-temperature samplers are able to explore a larger region of parameter space, allowing the sampler to transition between different modes, while swaps between the states of different rungs on the ladder allow the lower-temperature samplers to benefit from the increased exploration at higher temperatures.

In our problem, parallel tempering performs badly beyond the initial iteration (without coupling between nearby sightlines), as the samples we store from previous iterations are a poor representation of the higher-temperature distributions. We build a ladder of samplers, but employ a different parameterization to modify the target distribution. Our approach is to modify the inverse covariance matrix, in order to introduce correlations between voxels in neighboring distance bins. Instead of a temperature, we introduce a distance-mixing parameter, $\epsilon$, which allows us to smoothly interpolate between the approximately isotropic target distribution and distributions in which correlations in the radial direction are stronger than in the transverse direction.

The form of the deformation of the inverse covariance matrix is sketched out in Fig. \ref{fig:icov-deformation}. The entries linking neighboring distance bins of \textit{different} angular pixels increase with the distance-mixing parameter, $\epsilon$, according to the formula given in Fig. \ref{fig:icov-deformation}. The entries linking neighboring distance bins of the \textit{same} angular pixel (i.e., sightline) are fixed to zero. This ensures that when we sample a single sightline, keeping the neighbors fixed, the different distance bins are independent, and thus speeds up the sampling procedure significantly.

A the distance-mixing parameter of $\epsilon = 0$ indicates no change to the inverse covariance matrix. As $\epsilon$ increases, the penalty for moving a dust cloud forward or backward in a single distance bin typically decreases.

Just as in parallel tempering, we can treat the ladder of samplers as if they are sampling a single, augmented parameter space. If the original parameter space is denoted by $\theta$, then the augmented parameter space is the Cartesian product of $\theta_0 \times \theta_1 \times \theta_2 \times \cdots \times \theta_N$, with $\theta_i$ corresponding to distance-mixing parameter $\epsilon_i$, and $N$ being the number of rungs in the ladder. The probability density that we assign to the augmented space is given by
\begin{align}
    p \arg{\theta_0, \, \theta_1, \, \theta_2, \, \ldots, \, \theta_N}
    &= p_0 \arg{\theta_0} p_1 \arg{\theta_1} p_2 \arg{\theta_2} \cdots p_N \arg{\theta_N} \, ,
\end{align}
where $p_i \arg{\theta_i}$ uses the inverse covariance matrix deformed with parameter $\epsilon_i$. Because the probability density of the augmented space is factorizable in this manner, we can sample in each subspace independently. However, we introduce swap steps between neighboring rungs of the ladder, in which we propose to exchange the values of $\theta_i$ and $\theta_{i+1}$. The Metropolis-Hastings acceptance probability for this step is
\begin{align}
    \min \left[ \,
        \frac{
            p_i \arg{\theta_{i+1}} p_{i+1} \arg{\theta_i}
        }{
            p_i \arg{\theta_i} p_{i+1} \arg{\theta_{i+1}}
        } \, , \ 
        1
    \, \right] \, .
\end{align}
For parallel tempering, this acceptance probability simplifies down to an elegant form, but in the more general case of a ladder of modified probability densities, the above acceptance ratio does not necessarily simplify.

Notice that the cross-distance coupling introduced in the higher rungs of the ``deformation ladder'' only affects the Gaussian process prior that couples nearby sightlines. In the initial iteration, however, we treat each sightline independently, so the ``deformation ladder'' would have no effect. In the initial iteration, we therefore use standard parallel tempering in order to improve convergence of our Markov Chain Monte Carlo sampler. In correlated iterations, we use a ladder with three rungs, with the distance-mixing parameter increasing linearly from $\epsilon = \exp \arg{-2} \approx 0.14$ in the bottum rung to $\epsilon = \exp \arg{-\nicefrac{1}{4}} \approx 0.78$ in the top rung. Note that the distance-mixing parameter is non-zero in the bottom rung, meaning that our target posterior density is slightly modified, and contains small cross-distance correlations.

\section{Assessing convergence of the line-of-sight reddening sampler}
\label{app:convergence}

\begin{figure*}
    \centering
    \includegraphics[width=\textwidth]{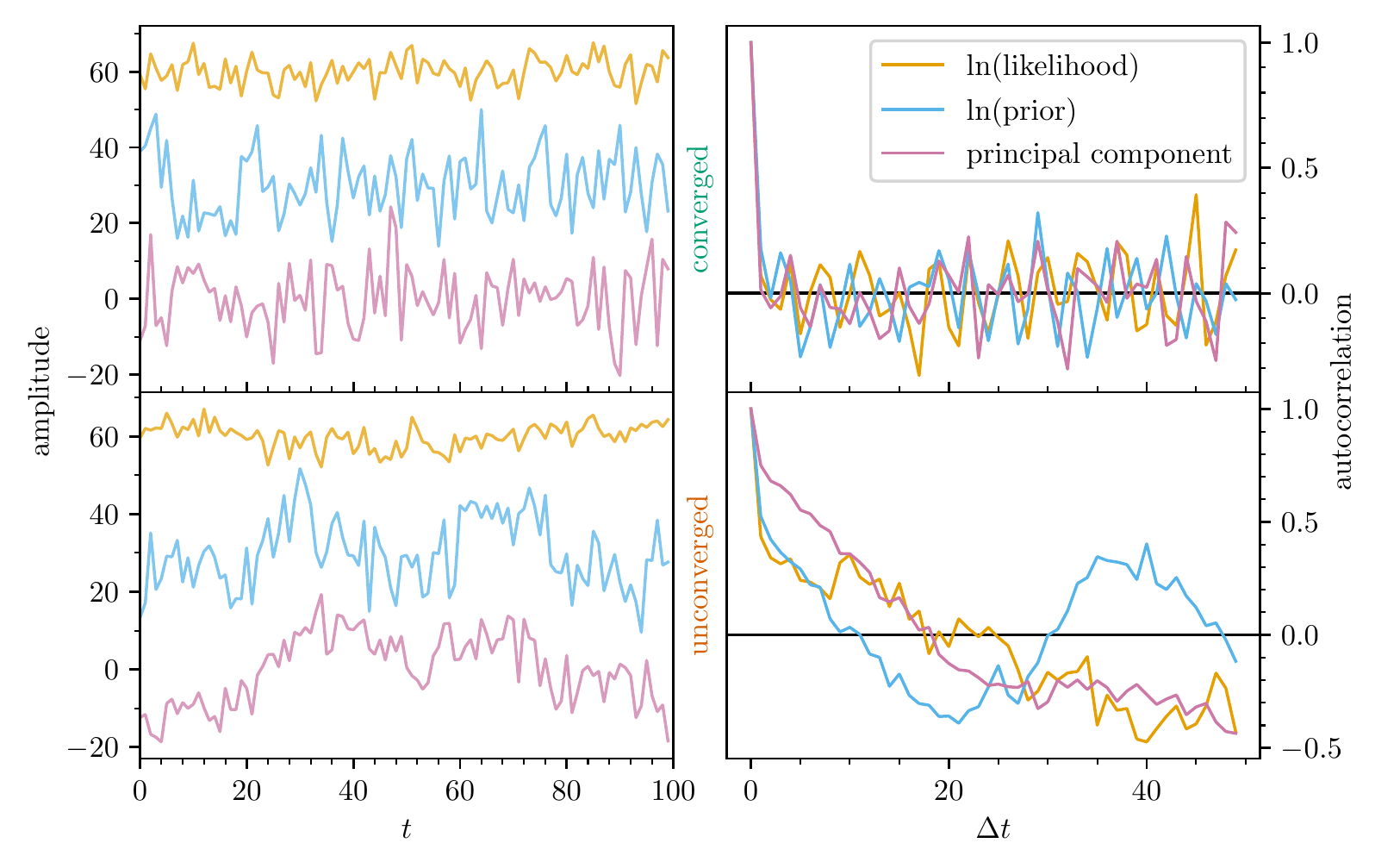}
    \caption{An example of our convergence diagnostic, for an MCMC run with good convergence (top panels) and with poor convergence (bottom panels). The left panels show three functions of the sampled parameters (the logarithms of the prior and likelihood, and the amplitude of the first principal component) as a function of sampling step. The right panels show the autocorrelation of these three functions. Non-convergence can be diagnosed from the long autocorrelation time in the bottom-right panel.}
    \label{fig:convergence-diagnostic}
\end{figure*}

We test for convergence of our line-of-sight sampling by measuring the autocorrelation time of certain functions of the line-of-sight reddening density parameters, $\alpha$. There are many different functions that one could choose to try to assess chain mixing, but we choose three: the prior, the likelihood, and the amplitude of the first principal component of $\alpha$ in the central pixel. This third function measures how quickly the sampler explores the direction in parameter space that accounts for most of the variance in the Markov Chain. Typically, this direction in parameter space corresponds to changing the distance to the dominant cloud along the line of sight. We require that the number of samples in the Markov chain be at least 20 times the autocorrelation time of each of these three functions. Fig. \ref{fig:convergence-diagnostic} illustrates these convergence criteria for both a converged and non-converged case.

For pixels that fail this convergence criterion, we sample again with more robust settings (i.e., more steps and additional rungs in the deformation ladder). If a pixel fails to converge within three attempts, we flag the pixel and continue. Fewer than 0.25\% of pixels fail this convergence test in any of the iterations.

\section{Fitting a line to data}
\label{app:linear-model}

Here, we describe our method of fitting a linear relation between two quantities with measurement uncertainties. In Section \ref{sec:stellar-reddenings}, we use this model to determine the relation between $A_V$, as measured by StarHorse, and either our inferred $\Egr$ or the inferred $\EBPRP$ from Andrae2018. The model we describe here is very similar to that described by \citet{HoggBovyLang2010}, with the addition of a systematic error floor in the measured values of $y$.

Our observed data consists of $N$ measured values of $\hat{x}$ and $\hat{y}$, with reported Gaussian uncertainties $\sigma_x$ and $\sigma_y$. We assume that each data point is either ``good,'' in which case the true values of $x$ and $y$ (without measurement error) follow a linear relation,
\begin{align}
    y = m x + b \, ,
\end{align}
or an ``outlier,'' in which case the true values of $x$ and $y$ are independently distributed Gaussian random variates, with means $\mu_{x, \, b}$ and $\mu_{y, \, b}$, and standard deviations $S_{x, \, b}$ and $S_{y, \, b}$. The prior probability of any given data point being an outlier is $P_b$. We additionally assume that the reported values of $\sigma_y$ are systematically underestimated, and that the true uncertainty in the measured value of $y$ is given by
\begin{align}
    \sigma_{y, \, \mathrm{true}}^2 = \sigma_y^2 + \sigma_0^2 \, .
\end{align}
In order to determine the calculate the likelihood of a single data point,
\begin{align}
    \mathcal{L} \arg{\hat{x} , \hat{y}}
    \equiv
    \pcond{
        \hat{x}, \hat{y}
    }{
        \sigma_x, \sigma_y, \sigma_0, m, b, P_b, \mu_{x,\,b}, S_{x,\,b}, \mu_{y,\,b}, S_{y,\,b}
    } \, ,
\end{align}
it is necessary to specify a prior on the true value of $x$. We impose a flat prior, which yields a closed-form solution for the likelihood, as long as the width of the prior is large compared to $\sigma_x$ and contains the measured value, $\hat{x}$. This unfortunately introduces one more arbitrary parameter into the problem, $\Delta x$, the width of the flat prior in $x$. We choose $\Delta x$ to be approximately the same as the range of observed $\hat{x}$, and the recovered parameters do not depend sensitively on the choice of $\Delta x$. It can be shown that the likelihood of a single data point in this model is given by
\begin{align}
    \mathcal{L} \arg{\hat{x} , \hat{y}}
    =& \,
    \frac{1 - P_b }{\Delta x} \ 
    \mathcal{N} \arg{
        \hat{y}
        \, | \,
        m \hat{x} + b , \,
        \sigma_y^2 + \sigma_0^2 + m^2 \sigma_x^2
    }
    \notag \\ & \ \ \ 
    + P_b \
    \mathcal{N} \left[
        \begin{pmatrix}\hat{x} \\ \hat{y}\end{pmatrix}
        \, \Biggr| \,
        \begin{pmatrix}\mu_{x,\,b} \\ \mu_{y,\,b}\end{pmatrix} , \,
        \begin{pmatrix}
            S_{x,\,b}^2+\sigma_x^2 & 0 \\
            0 & S_{y,\,b}^2+\sigma_y^2+\sigma_0^2
        \end{pmatrix}
    \right] \, .
\end{align}
We assume that the data points have independently distributed errors, so that the likelihood of the entire data set is the product of the individual likelihoods.

\begin{deluxetable}{cl|cl}
    \tablecaption{Priors on linear model \label{tab:linear-model-priors}}
    \tablehead{
        \colhead{Parameter} & \colhead{Prior} &
        \colhead{Parameter} & \colhead{Prior}
    }
    \startdata
    $m$ & $U \arg{\nicefrac{1}{6}, \nicefrac{3}{2}}$ & $b$ & $U \arg{-0.25, 0.25}$ \\
    $P_b$ & $U \arg{0,1}$ & $\ln \sigma_0$ & $U \arg{-8, 0}$ \\
    $\mu_{x,\,b}$ & $U \arg{0, 10}$ & $S_{x,\,b}$ & $U \arg{0.1, 10}$ \\
    $\mu_{y,\,b}$ & $U \arg{0, 3}$ & $S_{y,\,b}$ & $U \arg{0.1, 3}$ \\[0.25em]
    \enddata
\end{deluxetable}

In Section \ref{sec:stellar-reddenings}, we treat the StarHorse estimates of $A_V$ as $x$, and our $\Egr$ estimates or Andrae2018's $\EBPRP$ estimates as $y$. The priors on our model parameters are given in Table \ref{tab:linear-model-priors}. We use the \texttt{emcee} Python package \citep{ForemanMackey2013emcee} to sample from our model, using 90 walkers, 2500 steps per walker for burn-in, and subsequently 5000 steps per walker. The resulting median inferred values of $m$, $b$ and $\sigma_0$ are given in Section \ref{sec:stellar-reddenings}.

\bibliographystyle{aasjournal}
\bibliography{bayestar2019}



\end{document}